\documentclass[11pt, prb]{revtex4-1}

\usepackage{graphicx,rotating,subfigure}
\usepackage{amsmath}
\usepackage{pstricks}
\renewcommand{\tensor}[1]{\mathsf{#1}}
\renewcommand{\vec}[1]{\mathbf{#1}}

\begin{document}

\title{Cooee bitumen. Chemical aging}

\author{Claire A. Lemarchand}
\email{clairel@ruc.dk}
\affiliation{
  DNRF Centre ``Glass and Time'', IMFUFA,
  Department of Sciences, 
  Roskilde University, Postbox 260,
  DK-4000 Roskilde, Denmark $\mbox{}$
}

\author{Thomas B. Schr{\o}der}
\affiliation{
  DNRF Centre ``Glass and Time'', IMFUFA,
  Department of Sciences,
  Roskilde University, Postbox 260,
  DK-4000 Roskilde, Denmark $\mbox{} $ $\mbox{} \mbox{}$
}

\author{Jeppe C. Dyre}
\affiliation{
  DNRF Centre ``Glass and Time'', IMFUFA,
  Department of Sciences, 
  Roskilde University, Postbox 260,
  DK-4000 Roskilde, Denmark $\mbox{} $ $\mbox{}$
}

\author{Jesper S. Hansen}
\affiliation{
  DNRF Centre ``Glass and Time'', IMFUFA,
  Department of Sciences,
  Roskilde University, Postbox 260,
  DK-4000 Roskilde, Denmark
}

\begin{abstract}
We study chemical aging in "Cooee bitumen" using molecular dynamic simulations.
This model bitumen is composed of four realistic molecule types: saturated hydrocarbon, resinous oil,
resin, and asphaltene. The aging reaction is modelled by the chemical reaction: 
"2 resins $\rightarrow$ 1 asphaltene". Molecular dynamic simulations of four bitumen compositions,
obtained by a repeated application of the aging reaction, are performed. 
The stress autocorrelation function, the fluid structure, the rotational dynamics of the plane aromatic molecules,
and the diffusivity of each molecule,
are determined for the four different compositions.
The aging reaction causes a significant
dynamics slowdown, which is correlated to the aggregation of asphaltene molecules in larger
and dynamically slower nanoaggregates. Finally, a detailed description of the role of each molecule
types in the aggregation and aging processes is given.
\end{abstract}

\maketitle

\noindent Corresponding author: C. A. Lemarchand, E-mail: clairel@ruc.dk

\newpage \setlength{\baselineskip}{0.75cm}

\section{Introduction}
Road deterioration is partly due to the aging of bitumen, of which viscosity~[\onlinecite{shrp368, herrington, lu, ec140}]
and brittleness~[\onlinecite{petersen74},\onlinecite{wu2009}] increase over years.
Bitumen chemical species can be categorized into the SARA classification~[\onlinecite{astm}],
for which S stands for saturates, the first A for aromatics, R for resins, and the second A for asphaltenes.
Asphaltene molecules represent the crude oil fraction not soluble in alkane, usually heptane~[\onlinecite{astm, lesueur09, pierre04}] and
are the largest molecules found in crude oil.
Chemical aging is characterized by an increase of the asphaltene content in bitumen~[\onlinecite{shrp368, herrington, lu, ec140}].
Experimental work has shown that an increase of the asphaltene content leads to an increase in
viscosity~[\onlinecite{sheu, sirota, hasan}].

It is known from the experimental~[\onlinecite{yen, mullins2011, eyssautier2012, mullins2012}]
and theoretical~[\onlinecite{zhangFirst},\onlinecite{us}] literature that asphaltene molecules
tend to aggregate. In an aggregate the plane aromatic asphaltene bodies tend to align~[\onlinecite{zhangFirst},\onlinecite{zhang2008}]. 
We conjecture that nanoaggregates play a key role in the bitumen dynamics and lie behind the increase in viscosity observed as bitumen ages.
The role of the other molecule types described in the SARA classification in the asphaltene aggregation
is not completely clear~[\onlinecite{mullins2010}].

In this article, we model chemical aging through a reaction transforming resins into asphaltenes.
This reaction is supported by experimental evidence showing that
as bitumen ages, the average molecular mass increases~[\onlinecite{ec140, shrp368, lu, mouazen13}].
A model bitumen developed in a previous work~[\onlinecite{us}], in the framework of the Cooee project~[\onlinecite{cooee}],
is well adapted to implement
this aging reaction because it contains one molecule of each class described in the SARA classification.
In this context, we wish to address the following three questions:
\begin{enumerate}
\item Is there an effect of the aging reaction on the rheological properties?
\item If there is, is it correlated to asphaltene aggregation? 
\item What is the role of the other molecular species in the aggregation and in the aging processes?
\end{enumerate}

To answer these three questions, we performed molecular dynamic simulations of bitumen at four different chemical compositions,
representing different stages in the aging process.
For each composition, the rheological properties were first studied through the stress autocorrelation function.
Then, the aggregate structure was identified and quantified.
Finally, dynamical properties defined at the molecular level such as
the rotational dynamics of the aromatic plane molecules and the diffusivity of each molecule type, were computed for each composition.
They bring fundamental new insight about the aggregates' dynamics as bitumen ages and the role of each molecule in the aging and
aggregation processes.

The paper is organized as follows. Section~\ref{sec:method} contains numerical methods.
The stress autocorrelation function is studied in section~\ref{sec:stress}.
Section~\ref{sec:structure} quantifies the nanoaggregate structure.
Section~\ref{sec:rot} and~\ref{sec:diff} discuss results about rotational dynamics and diffusivity,
respectively. A model describing the role of each molecule in
the aggregation and aging processes is built up from these results.
Section~\ref{sec:conclu} contains a summary.

\section{Simulation method}
\label{sec:method}

\subsection{Molecular dynamics}
To study chemical aging of bitumen, we used molecular dynamic simulations, which have been employed previously to investigate
bitumen~[\onlinecite{zhangFirst, zhang2008, us, li2013}]. The simulations features used are described in detail in Ref.~[\onlinecite{us}]
and summarized here.

\subsubsection{GPU computing}
To probe the slow dynamics in bitumen, molecular dynamic simulations need to reach long time spans. To this end,
the simulations were performed
on Graphic-Processor-Units (GPU) using the RUMD package (version 1.2)~[\onlinecite{rumd}]. 
Each simulation performed in this work has a time span of $0.35$ $\mu$s, which corresponds to two weeks of computation time.

\subsubsection{Molecules}
\begin{figure}
  \scalebox{0.20}{\includegraphics{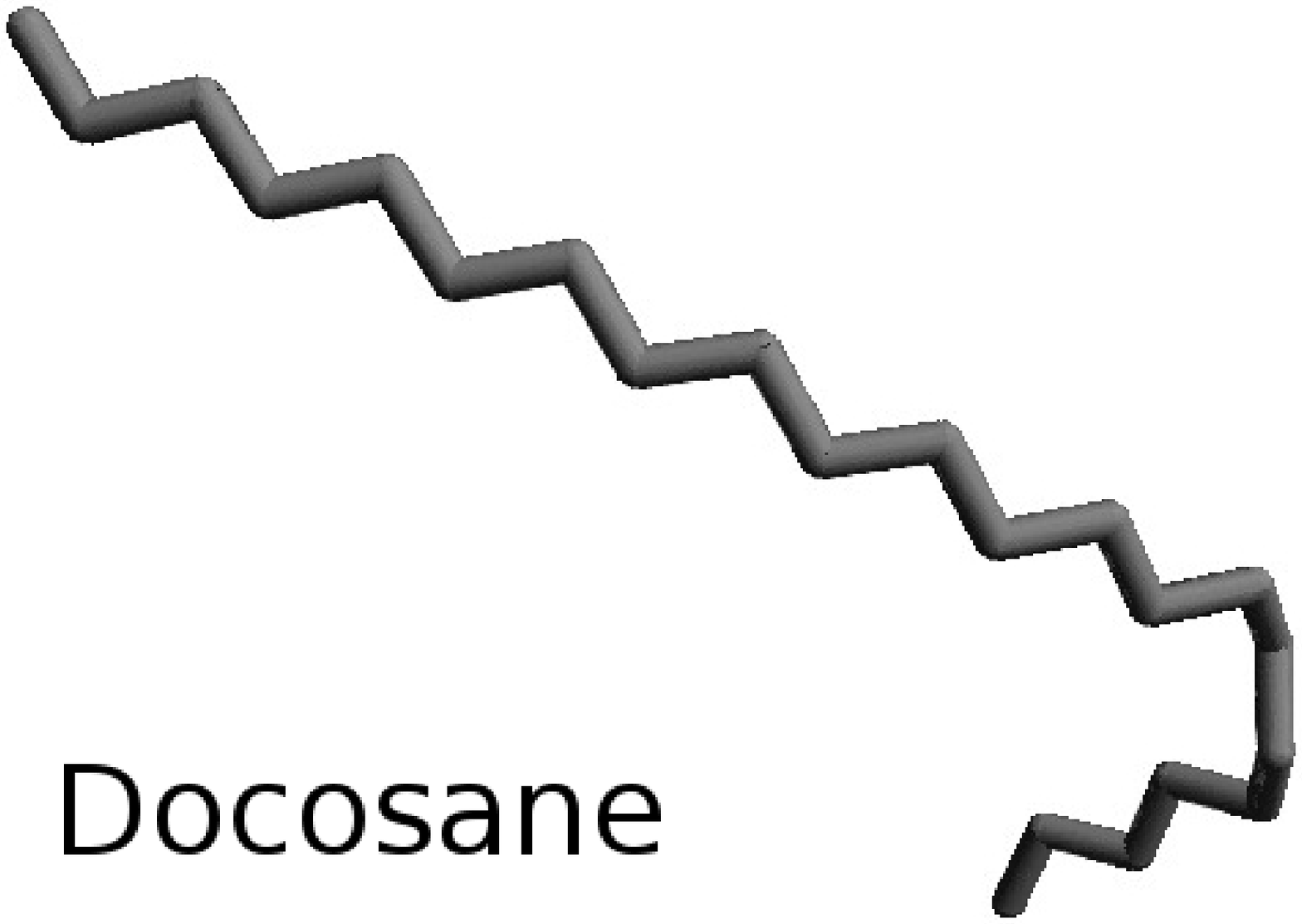}}
  \scalebox{0.20}{\includegraphics{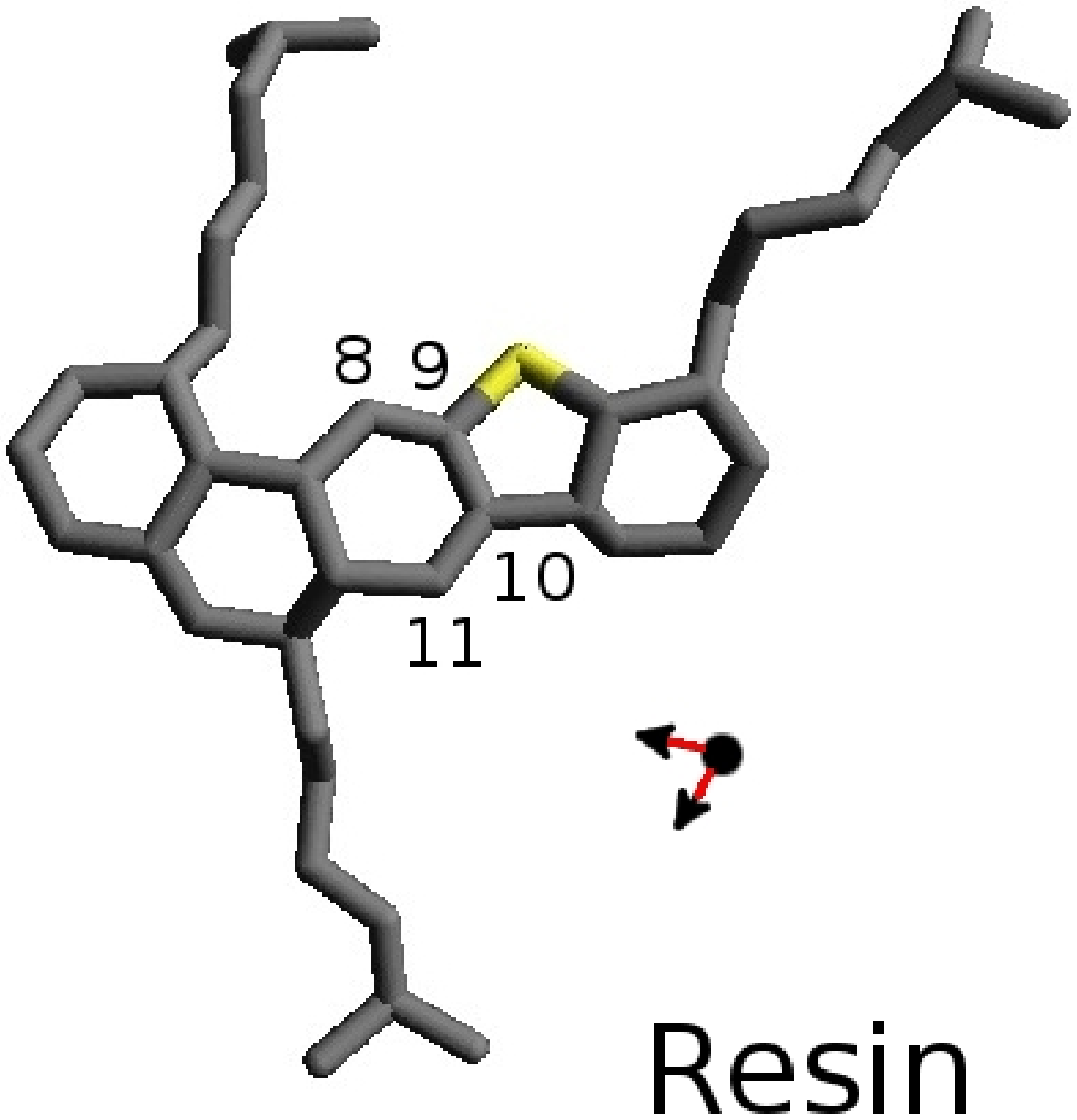}}\\
  \scalebox{0.25}{\includegraphics{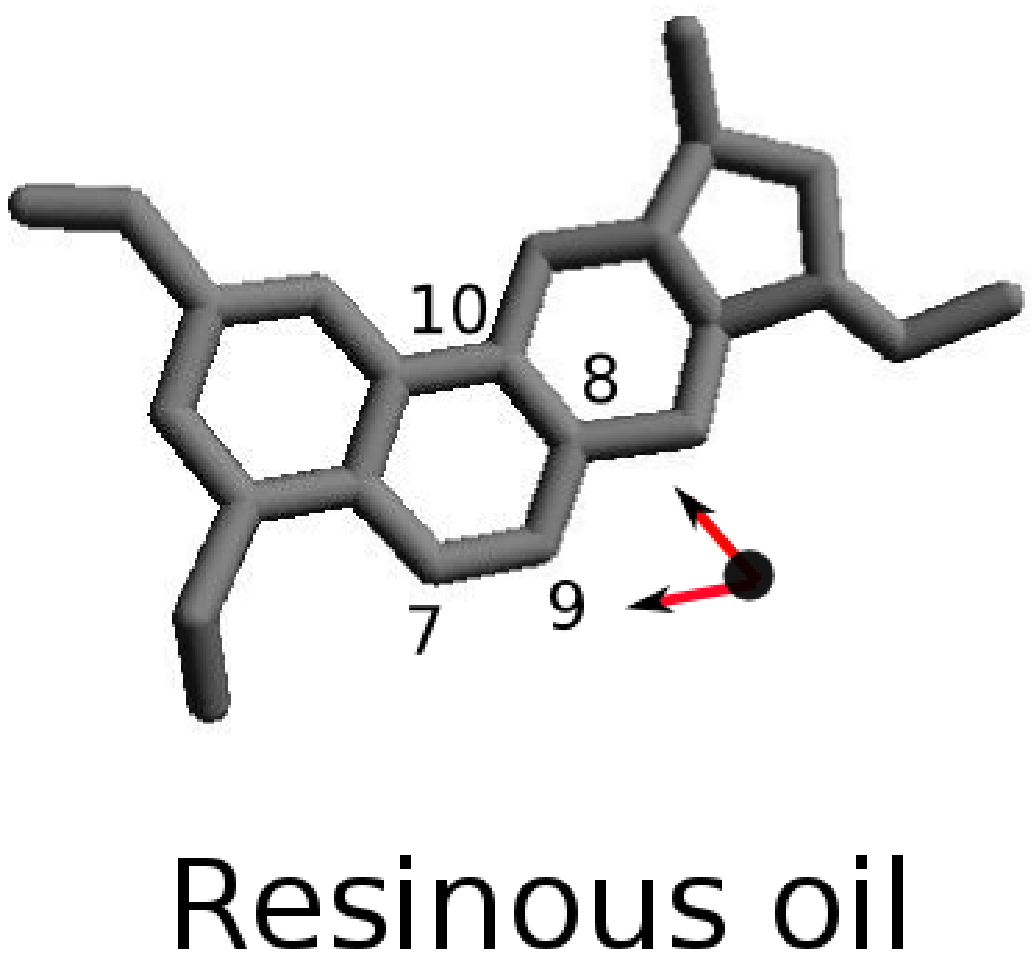}}
  \scalebox{0.25}{\includegraphics{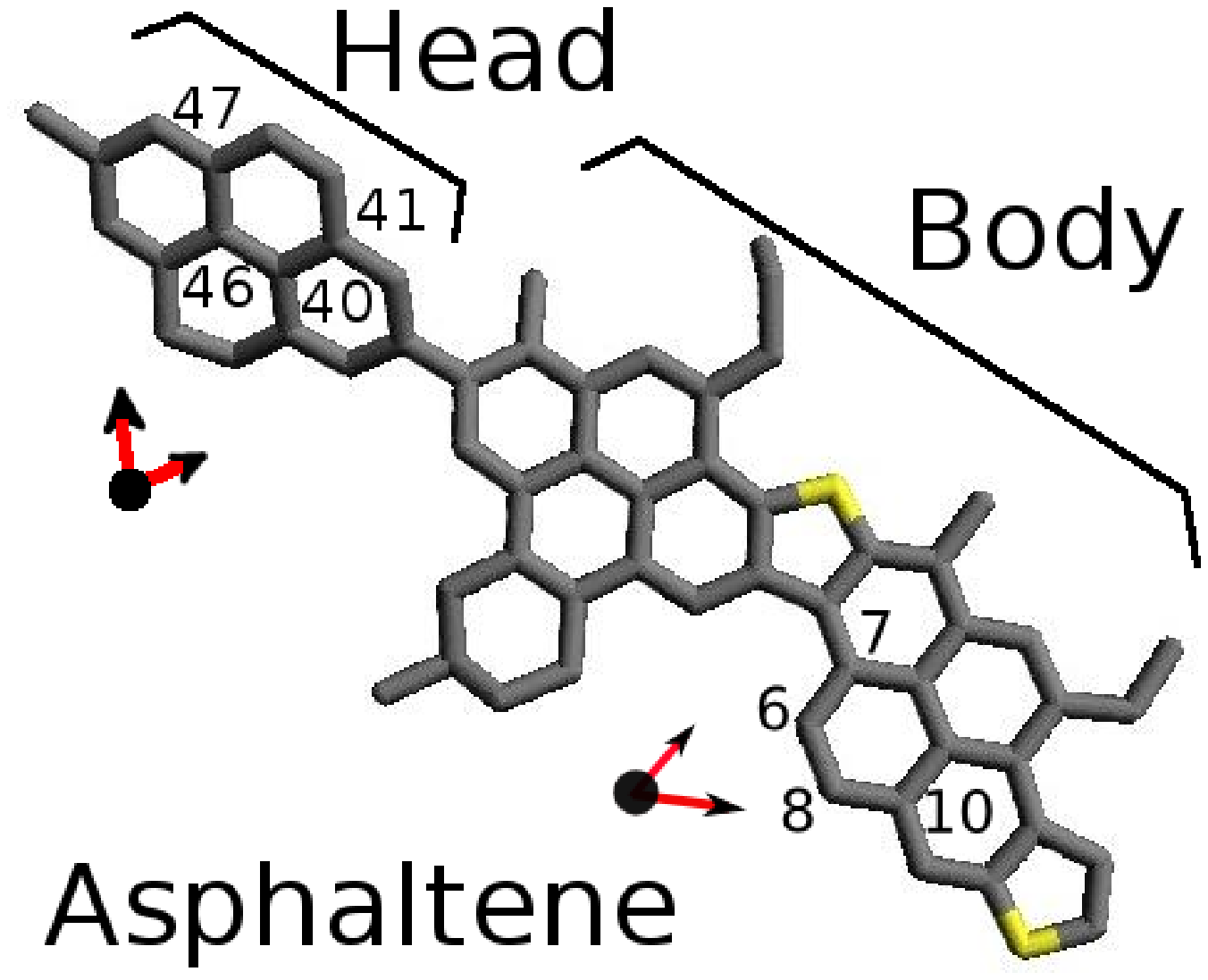}}
  \caption{\label{fig:molecule}
(Color online) Structure of the four molecules in the "Cooee bitumen" model. Grey edges represent
the carbon groups $CH_3$, $CH_2$ and $CH$ and yellow edges represent sulfur atoms.
The "head" and "body" of the asphaltene molecule are shown.
Numbers and arrows indicate bond-vectors used to quantify the nanoaggregates' structure and dynamics. See sections~\ref{sec:structure}
and~\ref{sec:rot} for details.
}

\end{figure}

To reproduce the SARA classification mentioned in the introduction, the Cooee bitumen contains four different molecule types:
asphaltene, resin, resinous oil which represents the aromatics class and docosane which falls into the saturates class~[\onlinecite{us}].
The structure chosen for each molecule is displayed in Fig.~\ref{fig:molecule}. 
The molecule structures come from the literature, see Refs.~[\onlinecite{artok, murgich, rossini, storm}].
The asphaltene molecule, chosen in this work, is the largest of the four molecules considered and contains two sulfur atoms. 
The resin molecule contains one sulfur atom.
All molecules are aromatic except for the docosane molecule.

\subsubsection{Interaction nature}
The methyl ($CH_3$), methylene ($CH_2$), and methine ($CH$) groups are represented by the same Lennard-Jones particle, a united atomic unit
(UAU) with molar mass 13.3 g.mol$^{-1}$ and Lennard-Jones parameters $\sigma = 3.75$ \AA $ $ and $\epsilon/k_B = 75.4$ K, where $k_B$ is the Boltzmann constant.
For simplicity, sulfur atoms are modelled by the same Lennard-Jones interaction,
but have a different molar mass of 32 g$\cdot$mol$^{-1}$. Thus in this description, molecules do not possess any polarity,
and intermolecular interactions are solely defined as van der Waals interactions.

Intramolecular interactions are modelled by flexible bonds with length $r_{ij}$ between two bonded UAUs $i$ and $j$,
angles $\theta$ between three bonded UAUs and dihedral angles $\phi$ between four bonded UAUs. The
consequent full force field is given by:
\begin{equation}
\begin{aligned}
  U(\vec{r})&=   \sum_{i} \sum_{j>i} 4\epsilon\left[
          \left(\frac{\sigma}{r_{ij}}\right)^{12} -
          \left(\frac{\sigma}{r_{ij}}\right)^{6}
          \right] 
          +\frac{1}{2}\sum_{\text{\tiny{bonds}}}k_s(r_{ij}-l_{\text{\tiny{b}}})^2\\
  &+\frac{1}{2}\sum_{\text{\tiny{angles}}}k_{\theta}(\cos \theta - \cos
  \theta_0)^2
  +\sum_{\text{\tiny{dihedrals}}}\sum_{n=0}^5 c_n\cos^n \phi.
\end{aligned}
\label{eq:forcefield}
\end{equation}
The values of the parameter $k_s$, $l_b$, $k_{\theta}$, $\theta_0$, and $c_n$ are listed in table~\ref{table:param},
taken from a previous work~[\onlinecite{us}].

\begin{table} [h]
  \begin{tabular}{lcc}
    \hline \hline
    Bonds                   & $l_b$  & $k_s$
    \\
    & [\AA]   & [kcal/mol]
    \\
    \hline
    All & 1.46     & 403 \\
    \hline \hline
    Angles                        & $\theta_0$      & $k_\theta$  \\
    & [Degrees]  & [kcal/(mol rad$^2$)]  \\
    \hline
    Aromatic/cyclo            &    120      & 108 \\
    Aliphatic           &    106      & 70 \\
    \hline \hline
    Dihedrals                   & Angle       &  $c_n$   \\
    &  [Degrees]  &   [kcal/mol] \\
    \hline
    Aromatic/cyclo  & 180  &  20  (n=1)\\
    Aromatic/cyclo  &   0  & -20  (n=1)\\
    Linear/aliphatic &   &  Ryckaert-Belleman \\
    \hline \hline
  \end{tabular}
  \caption{\label{table:param}
    Force-field parameters, see Eq. (\ref{eq:forcefield}). From Ref.~[\onlinecite{us}].
  }
\end{table}

\subsubsection{Choice of temperature, pressure and density}
Simulations are performed in the canonical ensemble (NVT with a Nos\'e-Hoover thermostat) at a constant temperature $T = 452$ K and a constant density.
The density is scaled in each simulation so that the average pressure of the system is around $1$ atm.
The temperature chosen $T = 452$ K is close to the temperature at which bitumen and stones are processed and mixed.
It is also a high enough temperature so that the slow bitumen dynamics can be probed using molecular dynamics~[\onlinecite{us}]. 

\subsection{Modeling chemical aging}
In order to take chemical aging into account in the molecular dynamic simulations, we chose to simulate four systems with
four different chemical compositions. To go from one composition to another the reaction "2 resins $\rightarrow$ 1 asphaltene"
is applied. 
This reaction is chosen because some experimental results suggest that as bitumen ages, the asphaltene content increases and
the resin content undergoes the greatest mass loss~[\onlinecite{ec140}].
However, the experimental literature proposes also other scenarios for chemical aging:
the aromatic content could decrease while the resin and asphaltene contents increase~[\onlinecite{lu}] or the resin and aromatic contents
increase due to a decrease in the saturates content while the asphaltene content is mainly constant but composed of molecules
with increasing molecular mass~[\onlinecite{mouazen13}]. In every cases, chemical aging is linked to an increase in the
average molecular mass.
The main effect of the aging reaction chosen in this work is to fuse molecules
together, resulting also in an increase of the average molecular mass.
For the sake of simplicity, we considered only one aging reaction, satisfying the basic principle of increasing molecular mass.
Total mass is not strictly constant in this reaction but vary very little. 
The exact composition of the four mixtures is given in table~\ref{tab:compo}.
The number of molecules of each type was implemented in each mixture and the corresponding weight fractions were calculated
knowing the molar mass of each molecule.
The weight fractions of asphaltene molecules in mixtures I and IV are close to the weight fractions of asphaltene
molecules in an unaged AAD-1 bitumen and in the same bitumen aged through the TFAAT (Thin Film Accelerated Aging Test) method
at 113$^\circ$C for 72 hours, respectively~[\onlinecite{shrp368}, \onlinecite{ec140}]. The natural abundance in asphaltene molecules
in AAD-1 bitumen is around 20\% and it reaches 36\% after aging through TFAAT. The TFAAT method is known to
reproduce chemical aging of bitumen for a period between 5 to 10 years~[\onlinecite{shrp368}, \onlinecite{ec140}]. This value is our
best estimate of the aging period our reaction models.

\begin{table}
\begin{center}
\begin{tabular}{ccccccc}
\hline
 mixture  &$\rho$ (kg.m$^{-3}$) & A(\%)  & R(\%) & RO(\%) & D(\%) \\
\hline \hline
 I    & 964 & 10 (21.5)  &  10 (13.8)      & 10 (7.6)   & 82 (57.1) \\
\hline
 II   & 967 & 13 (28.5)  & 4 (5.6)      & 10 (7.7)   & 82 (58.2) \\
\hline
 III  & 969 & 14 (30.8)  & 2 (2.8)      & 10 (7.8)  & 82 (58.6)\\
\hline
 IV   & 973 & 15 (33.2)  &  0      & 10 (7.9)  & 82 (58.9) \\
\hline
\end{tabular}
\end{center}
\caption{Density $\rho$ and number of asphaltene (A), resin (R), resinous oil (RO), and docosane (D) molecules
in the four different mixtures. The numbers in brackets refer to the weight fraction.
For each mixture the temperature is $T = 452$ K and the average pressure is $P = 1 \pm 250$ atm.
}
\label{tab:compo}
\end{table}

To obtain statistically reliable results eight to ten simulations were carried out for each chemical composition,
starting from independent initial conditions. 
Each simulation was first equilibrated for $17$ ns before results were computed. 

\section{Stress autocorrelation function}
\label{sec:stress}

The first question, which is interesting from the point of view of bitumen applications is: "Is there an effect
of the aging reaction on bitumen's rheological properties?".
As a first attempt to quantify the rheological properties, we studied the stress autocorrelation function.

The pressure tensor $\tensor{P}$ is defined from molecular quantities as in our previous work~[\onlinecite{us}] using the Irving-Kirkwood expression~[\onlinecite{irving}]:
\begin{equation}
\label{stress}
\tensor{P} = \frac{1}{V} \sum_{i=1}^{N} \Bigl [  \frac{\mathbf{p}_i\mathbf{p}_i}{m_i} + 
\sum_{j=i+1}^N  \mathbf{r}_{ij} \mathbf{F}_{ij} \Bigr ] ,
\end{equation}
where $V$ is the system volume, $\mathbf{p}_i$ the momentum of molecule $i$, $m_i$ the molecule mass,
$\mathbf{r}_{ij} = \mathbf{r}_{i} - \mathbf{r}_{j}$ the difference between the center-of-mass position of molecules $j$
and $i$ and $\mathbf{F}_{ij}$ is the total force applied by molecule $i$ on molecule $j$.
The total force between two molecules is computed as the sum of the atomic forces acting between the atoms of the two molecules.
The traceless symmetric part $\stackrel{\mathrm{os}}{\tensor{P}}$ of the pressure tensor defined in Eq.~\eqref{stress}
can be written as
\begin{equation}
\stackrel{\mathrm{os}}{\tensor{P}} = \frac{1}{2}(\tensor{P} + 
\tensor{P}^T) - \frac{1}{3}\mathrm{trace}(\tensor{P}) \tensor{I},
\end{equation}
where the superscript $T$ denotes the transpose and $\tensor{I}$ is the identity tensor.
The shear stress autocorrelation function~$C(t)$ is then defined as:
\begin{equation}
\label{stressACF}
C(t) =  \frac{1}{3}\sum_{(\alpha \beta)}
\Bigl \langle \stackrel{\mathrm{os}}{\tensor{P}}_{(\alpha \beta)}(0)
\stackrel{\mathrm{os}}{\tensor{P}}_{(\alpha \beta)}(t) \Bigr \rangle \ ,
\end{equation}
where $t$ is time, $(\alpha \beta)$ runs over the off-diagonal stress
tensor elements, and $\langle \cdot \rangle$ denotes an ensemble
average. For each independent simulation, the average is done over 100 consecutive intervals
of $3.5$ ns, the entire simulation being $0.35$ $\mu$s long. For mixture I, the computation of the
stress autocorrelation function is extended to a period of $35$ ns. The last decade corresponds to
an average over 10 consecutive intervals of the entire simulation.

The stress autocorrelation function is useful because its integral over time leads to the viscosity $\eta$
of the mixture through the Green-Kubo relation:
$\eta = (V/k_BT)\int_0^\infty C(t) dt$,
where $k_B$ is the Boltzmann factor. 
However, as will be shown in the following, the dynamics in bitumen is so slow that the stress
autocorrelation function $C(t)$ is not fully decayed even after the time $35$ ns accessible through molecular dynamic simulations.
Consequently, it is not possible to obtain a reliable value of the viscosity in molecular
dynamics through the Green-Kubo relation.
Nevertheless, some valuable knowledge can be obtained from the stress autocorrelation function.

\begin{figure}
\includegraphics[scale=0.4, angle=-90]{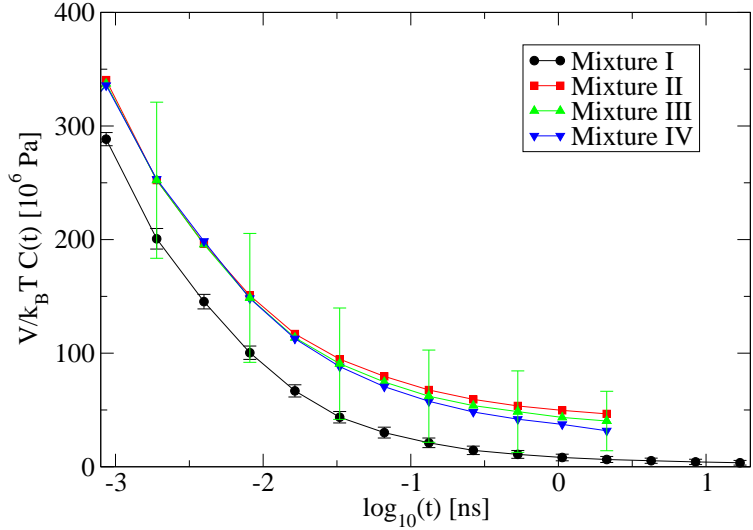}
\caption{
(Color online). Time evolution of the logarithmic average of the normalized stress autocorrelation function $\Bigl(V/k_BT\Bigr)$ $C(t)$ for
the four different mixtures.
Error bars correspond to the standard deviations computed over the ten independent simulations performed for mixture I
and over the nine independent simulations performed for mixture III.
The initial value of the stress autocorrelation function (not visible) is $(V/k_BT)$ $C(0) > 10^{10}$ Pa in each mixture.
}
\label{fig:sacfLogAvg}
\end{figure}

To compare the stress autocorrelation function decay in the different mixtures, we computed
its "logarithmic average", as defined below.
The data were collected linearly in time.
Thus, the number of data points at long times is very large. This excess of points can be used to reduce noise at longer times.
The logarithmic average is done over an exponentially increasing number of points as time goes by. Formally,
for a logarithmic average in base 2, one can write:
\begin{equation}
\bar{C}_j = \frac{1}{2^j-2^{j-1}} \sum_{i=2^{j-1}+1}^{2^j} C_i \quad \text{ with } j \in [1, \text{Int}(\log_2(N))],
\end{equation}
where $\bar{C}_j$ is
the logarithmic averaged stress autocorrelation function at a time index $j$ running exponentially
and $C_i$ the stress autocorrelation function at a time index $i$ running linearly, $N$
is the total number of points in the linear sequence, Int$(\log_2(N))$ is the integer part of the base 2 logarithm of $N$.
The same average is applied to time.
The logarithmic average of the stress autocorrelation was computed for each simulation of each mixture. 
A standard average over the different independent simulations was finally performed for each mixture.
The corresponding average stress autocorrelation function is plotted versus time in Fig.~\ref{fig:sacfLogAvg} for the four mixtures.
Figure~\ref{fig:sacfLogAvg} shows that, on average, the stress autocorrelation decay is much faster in mixture I, than in the other
three mixtures. Mixtures II to IV have similar behavior with respect to the stress autocorrelation decay.
The standard deviation computed over the different independent simulations is indicated for mixture I and III 
(it is not indicated for mixtures II and IV for clarity, as it is close to the one plotted for mixture III).
It is to be noticed that the standard deviation of the stress autocorrelation function
is much smaller in the unaged mixture than in the aged mixtures.
This is probably due to the following: in the aged mixture, the system
can be stuck in its initial configuration during time spans accessible to molecular dynamics. Consequently,
standard deviations computed from simulations with independent initial configurations can be very large.
These large standard deviations
are consistent with the dynamics slowdown of the stress autocorrelation function observed on average for the aged mixtures.
As the viscosity is defined as the time integral of the stress autocorrelation function,
Fig.~\ref{fig:sacfLogAvg} indicates that the viscosity in the aged mixtures might be a lot larger than the viscosity in the unaged mixture.
This is in agreement with the experimental literature on aged bitumen~[\onlinecite{shrp368, herrington, lu, ec140}].

At this point, it is worth expanding on the results obtained in the bitumen literature~[\onlinecite{shrp368, herrington, lu, ec140}].
Aging is often referred to as an oxidation process because it has been observed that the content of oxidized polar molecules
increases as bitumen ages.
It is not inconsistent with the increase of asphaltene molecules also observed as bitumen ages,
because asphaltene molecules are indeed polar.
It has been argued~[\onlinecite{shrp368, herrington, lu, ec140}] that the increase in viscosity observed
as bitumen ages is due to stronger interactions between more and more polar molecules.
In our model bitumen, the molecules do not possess any dipole moment
and intermolecular interactions are only described using the van der Waals potential.
This is sufficient to produce an increase in the stress autocorrelation function relaxation time, as bitumen ages.
It means that, even in real bitumen, van der Waals interactions 
alone can account for the observed increase in bulk viscosity.

Why does the stress autocorrelation function decay more slowly in an aged bitumen
than in a younger one?
We conjecture that this is linked 
to the aggregation of asphaltene molecules. The first step to prove that there is a correlation
between the two phenomena
is to identify the nanoaggregates' structure in the different mixtures. This is the aim of section~\ref{sec:structure}.

\section{Nanoaggregate structure}
\label{sec:structure}

The previous section showed that aged bitumen have a stress autocorrelation function that decays more slowly
than unaged bitumen. In order to investigate whether it is linked to the existence of asphaltene nanoaggregates
in the system, the first step is to observe the aggregates' structure in the different mixtures.
The second step is to quantify a change in the aggregates' structure in the different mixtures. It is done in this section firstly by studying
the radial distribution function and secondly by designing a precise definition of a nanoaggregate at
the molecular level.

\begin{figure}
\includegraphics[trim=1cm 4cm 3cm 4cm, clip =true, scale=0.7]{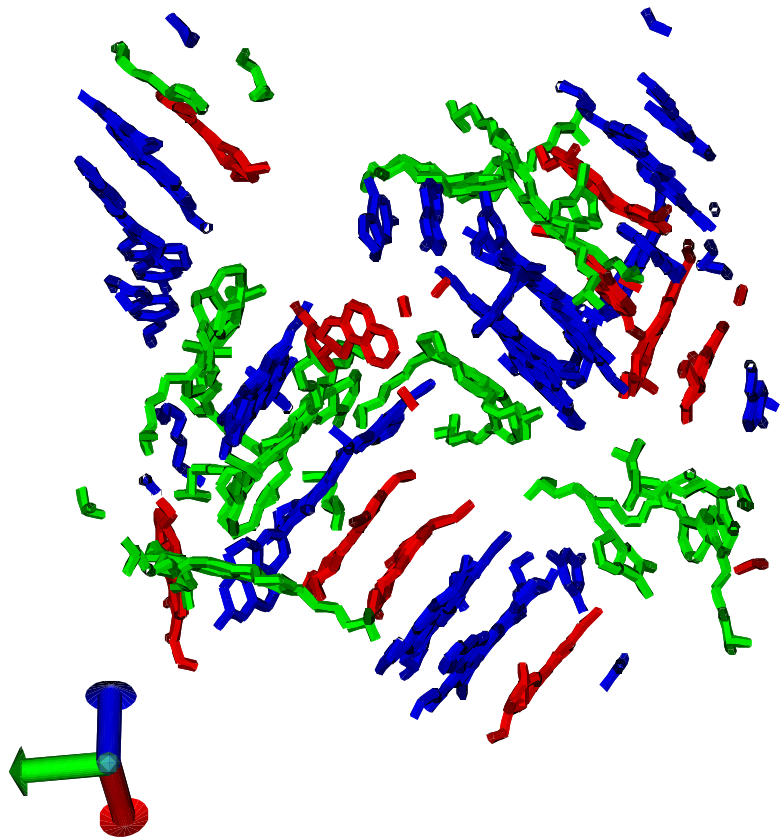}
\caption{
(Color online). Snapshot of molecule positions for mixture I.
Asphaltene molecules are colored in blue, resin in green and resinous oil in red. Docosane molecules are omitted for clarity.
}
\label{fig:aggregate}
\end{figure}

Figure~\ref{fig:aggregate} represents a snapshot of mixture I, where only asphaltene, resin and resinous oil molecules
are shown. It is clear from this figure that the asphaltene molecules align and form nanoaggregates.
Resin and resinous oil molecules
also have a flat aromatic structure and align with the asphaltene structure. 
An aligned geometry is preferred in our simulations because it minimizes the van der Waals interactions between
the flat molecules.
This kind
of interaction between flat aromatic molecules is also known as "$\pi$-stacking"~[\onlinecite{zhangFirst},\onlinecite{zhang2008}].
Docosane molecules are present around the aggregates (not shown),
but do not participate in the aggregate structure.

\begin{figure}
\includegraphics[angle=-90, scale=0.4]{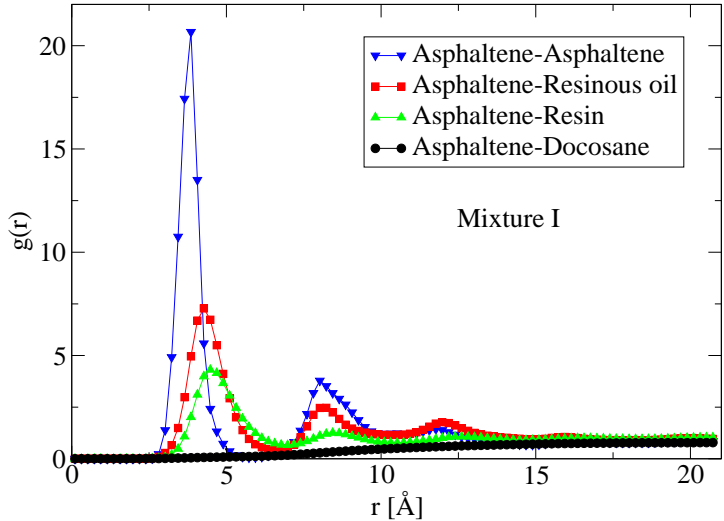}
  \caption{\label{fig:rdf}
    (Color online). The radial distribution function $g(r)$ for asphaltene-asphaltene (blue triangle down), asphaltene-resinous oil
(red square), asphaltene-resin (green triangle up), asphaltene-docosane (black circle) pairs in mixture I.
}

\end{figure}

To quantify the nanoaggregate structure,
the radial distribution function of the distance between asphaltenes, asphaltenes and resins, asphaltenes and
resinous oils and asphaltenes and docosanes can be plotted.
The radial distribution functions corresponding to the asphaltene-asphaltene, asphaltene-resin, asphaltene-resinous oil
and asphaltene-docosane pairs are plotted in Fig.~\ref{fig:rdf} for mixture I.
The distances were computed between the molecules' centers of mass.
The curves are obtained by averaging over the ten independent simulations of 0.35 $\mu$s each.

\begin{figure}
\includegraphics[scale=0.7]{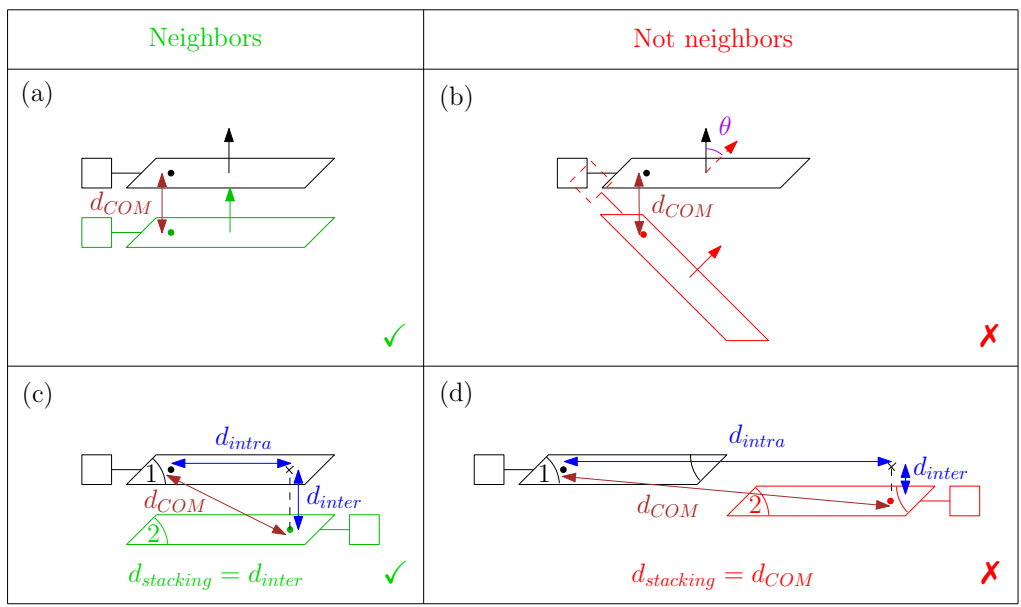}
\caption{
(Color online).
Scheme of four borderline cases illustrating whether an asphaltene molecule is or is not the nearest neighbor of another one.
The first asphaltene molecule is black. The second one is green if it is the nearest neighbor of the first one and red otherwise
(a) Head-to-head conformation. (b) Non-aligned molecules. (c) Head-to-tail conformation. (d) Molecules far from each other but aligned
in the same plane. $d_{\text{COM}}$ is the distance between the molecules' center of mass, $d_{\text{inter}}$ is the distance between
the two planes, $d_{\text{intra}}$ is the intraplanar distance, and $\theta$ is the alignment angle.
}
\label{fig:scheme}
\end{figure}

For the asphaltene-asphaltene pairs, a large peak is seen around 3.9 \AA,
reflecting the high probability of two asphaltene molecules to be perfectly aligned.
This distance is very close to the distance at which the Lennard-Jones potential describing the van der Waals interactions
between molecules has a minimum.
A second peak around $2\times3.9 = 7.8$ \AA$ $ can also be seen and reflects the extension of the stacked structure.
However, after the second peak the layered structure is quickly lost. This is partly due to the fact
that asphaltene molecules may
align in different ways. Asphaltene molecules are composed of a flat body and a flat head, which can rotate with respect to each other.
These two parts of the asphaltene molecules are shown in Fig.~\ref{fig:molecule}.
Bodies can be aligned in two different ways, either perfectly parallel, in a head-to-head conformation or in a head-to-tail conformation.
These two conformations are shown in Fig~\ref{fig:scheme} (a) and (c).
Only the first conformation leads to a distance between the centers of mass close to $3.9$ \AA.
The other conformation leads to bigger distances around $9$ \AA. 
The head of one molecule can also be aligned with the body of another molecule.
These different conformations were also observed by Zhang and Greenfield in Ref.~[\onlinecite{zhangFirst}].
A fourth conformation with only heads of the asphaltene molecules aligned was also identified in our simulations (not shown).
These different conformations induce an attenuation of the peaks in the radial distribution function as the distance between two
molecules increases.

For the asphaltene-resin and asphaltene-resinous oil pairs, a layered structure can also be inferred from the radial distribution
function. It follows the structure created by the asphaltene molecules. 
The first peak position of these two pairs is slightly
shifted to larger distances with respect to the first peak position of the asphaltene-asphaltene pairs.
This is probably due to the fact that resin and resinous oil molecules are smaller than asphaltene molecules, so that they can align
on top of an asphaltene molecule and glide over it at different distances from the asphaltene center of mass.
In the case of resin molecules, this effect can also be explained by the fact that the resin molecules
chosen in the Cooee bitumen model have side chains which can push
the molecules' center of mass away from the asphaltene plane.
In the case of asphaltene-resin and asphaltene-resinous oil pairs, the extension of the layered
structure is quickly smoothened out as the distance becomes larger. 

\begin{figure}
\includegraphics[angle=-90, scale=0.4]{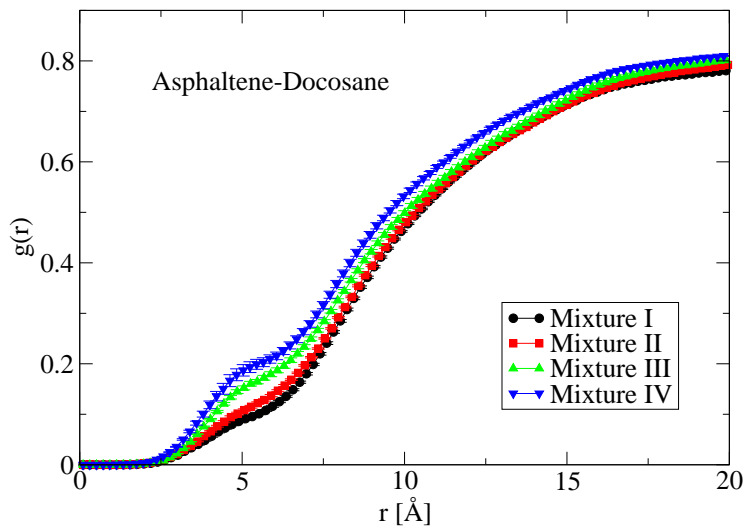}
\caption{
(Color online). Radial distribution function $g(r)$ for asphaltene-docosane pair in all mixtures.
Error bars correspond to the standard deviations computed over the 8 or more independent simulations performed in each case.
}
\label{fig:rdfDoco}
\end{figure}

The radial distribution function for the asphaltene-docosane pairs is represented for each mixture in Fig.~\ref{fig:rdfDoco}.
For the asphaltene-docosane pair the radial distribution function is very low around $3.9$ \AA $ $,
reflecting the fact that
docosane molecules are not members of the asphaltene nanoaggregates.
However, there is a slightly pronounced peak around $3.9$ \AA.
It can be seen from Fig.~\ref{fig:rdfDoco} that this peak
is growing monotonically from mixtures I to IV. It is probably due to the fact that as resin molecules disappear and asphaltene
molecules become more numerous, aggregates are mainly composed of asphaltene molecules. Thus, the edge of each aggregate is
likely to be represented by an asphaltene molecule next to a docosane molecule. 

In contrast to the asphaltene-docosane pairs, the standard deviations for the radial distribution functions
of the asphaltene-asphaltene, asphaltene-resin and asphaltene-resinous oil pairs are very large.
In other words, the radial distribution functions for these pairs can be very different from one simulation to another 
for the same mixture. For the sake of clarity, the standard deviations were not shown on the radial
distribution functions in Fig.~\ref{fig:rdf}.
The large standard deviation is due to the fact that one initial nanoaggregates' configuration
is preserved for a long time throughout a simulation and can be very different from the initial configuration
of another independent simulation.
The radial distribution function is consequently not the best way to quantify the nanoaggregate structure.

To deal with this problem, we propose to define a nanoaggregate at the molecular level using the following rule.
Two molecules are nearest neighbors and in the same nanoaggregate if:
\begin{enumerate}
\item their bodies are well-aligned AND close enough,
\item OR their heads are well-aligned AND close enough,
\item OR one head is well-aligned AND close enough to the body of the other molecule.
\end{enumerate}
Only asphaltene molecules can be characterized by a head and a body, whereas resin and resinous oil molecules only have a body.
The previous rule applies to them only when bodies are concerned. In other words,
resin and resinous oil molecules are treated as asphaltene bodies. Docosane molecules are not part of
a nanoaggregate.

\begin{figure}
  \scalebox{0.40}{\includegraphics[angle=-90]{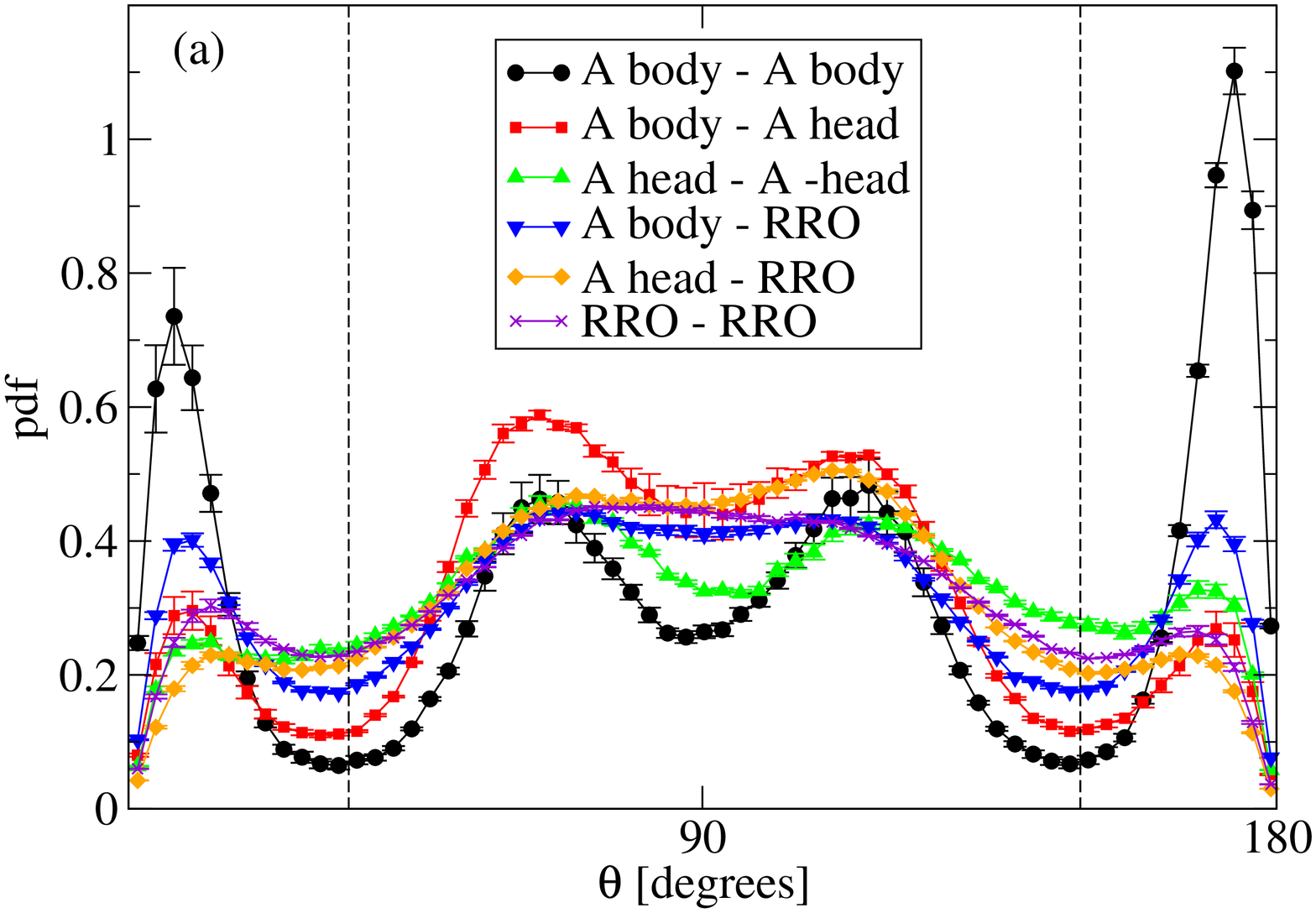}}\\
  \scalebox{0.40}{\includegraphics[angle=-90]{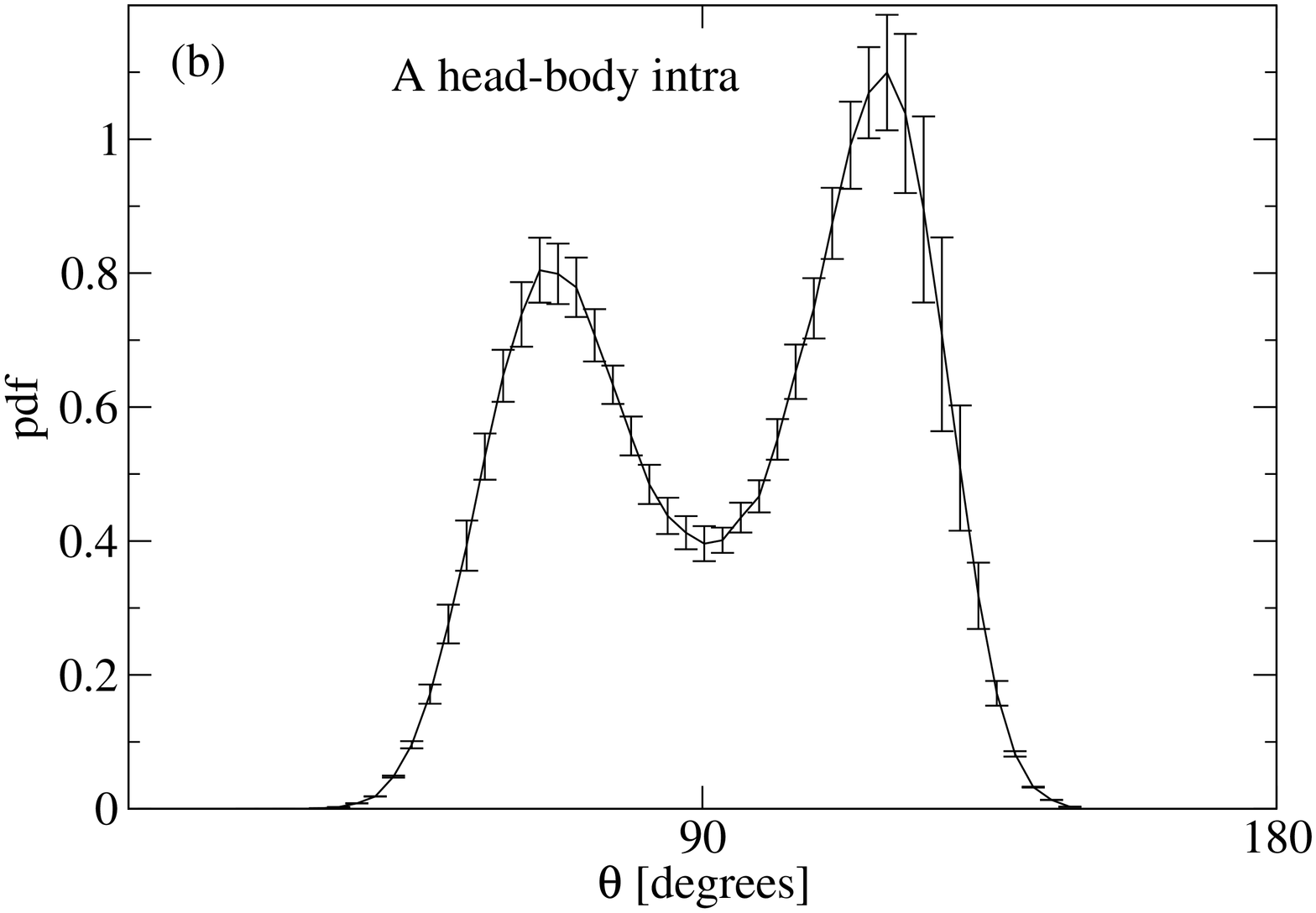}}
  \caption{\label{fig:pdfAngle}
(Color online) (a) Probability density functions (pdfs) of the intermolecular angle between
two asphaltene bodies (black circle), an asphaltene body and another asphaltene head (red square),
two asphaltene heads (green triangle up), an asphaltene body and a resin or resinous oil molecule (blue triangle down),
an asphaltene head and a resin or resinous oil molecule (orange diamond), and two resin or resinous oil molecules (purple cross),
in mixture I.
The dashed lines represent the threshold values chosen.
(b) Probability density function (pdf) of the intramolecular angle between the head and body of an asphaltene molecule
in mixture I. The error bars correspond to the ten independent
simulations performed for mixture I.
}

\end{figure}

The properties "well-aligned" and "close-enough" need to be quantified. 
Figure~\ref{fig:scheme} shows different cases illustrating which quantities and threshold values
are used to define "well-aligned" and "close-enough".
The alignment of
the plane aromatic molecules is quantified by the angle between the normal unit vectors
to each plane. 
In our previous work~[\onlinecite{us}], normal unit vectors were defined for
the asphaltene body and the resinous oil molecule. 
Similarly, the normal vector to the plane aromatic structure of the resin molecule is defined
as the cross product of two vectors pointing along a chemical bond. 
The normal vector to the asphaltene head is defined in the same way.
These four vectors
are shown in Fig.~\ref{fig:molecule}.
The dot product between the two normal unit vectors of two different molecules gives the cosine of the alignment angle between them.
Equivalently it provides us with an alignment angle between $0$ and $180$ degrees.
The probability density functions (pdfs) of the alignment angles between the different kinds of molecule are plotted in Fig.~\ref{fig:pdfAngle} (a).
The pdfs are computed for the alignment angle between any molecule of one type and any different molecule of the second type, whatever
the distance between them.
The pdfs have the same basic features. Each pdf possesses four peaks: one close to $0$ degree, two centered around $60$ and
$120$ degrees and a last one close to $180$ degrees. The two peaks close to $0$ and $180$ degrees correspond to molecules
aligned on top of each other either in a parallel or an antiparallel conformation.
To understand the existence of the two peaks centered around $60$ and
$120$ degrees, it is necessary to look at the probability density function of the angle between the head and the body of the
same asphaltene molecule. It is displayed in Fig.~\ref{fig:pdfAngle} (b). The pdf of the intramolecular angle
between the head and the body of an asphaltene molecule has two maxima at $60$ and $120$ degrees.
These values are monitored by the dihedral potential chosen in this work and correspond to the known conformation "gauche".
Inside a nanoaggregate, where two asphaltene bodies can be aligned in a head-to-tail conformation
without their heads to be aligned, this results in an angle of $60$ or $120$ degrees between the head of one asphaltene
molecule and the body of the other one, just as the intramolecular angle.
It explains the two peaks centered around $60$ and $120$ degrees observed
in the pdf of the angle between the body of an asphaltene molecule and the head of another asphaltene molecule.
The same explanation applies for the peaks centered around $60$ and $120$ degrees in the pdfs
of the intermolecular angle between asphaltene bodies and asphaltene heads.
For the pdfs of the angle between one resin/resinous oil molecule and another 
resin/resinous molecule or one asphaltene body or head, the peaks centered around $60$ and $120$ degrees are less well-defined
and can merge into one broad peak around $90$ degrees. They can be explained by the fact that in a nanoaggregate, if one resin or resinous oil molecule
is aligned with an asphaltene body, it has an angle of approximately $60$ or $120$ degrees with the same asphaltene head and the same angle
with another resin or resinous oil molecule aligned with the same asphaltene head. 
The peaks are broadened because they are due to angles between molecules which are not nearest neighbors.
To summarize, all the peaks observed around $60$ and $120$ degrees
in the probability density functions of the intermolecular angle can be explained by the intramolecular angle between the head
and body of the same asphaltene molecule. The asphaltene molecules thus impose their structure on the whole nanoaggregate,
which is to be expected as they are the largest molecules.

Important for the nanoaggregate definition are the peaks close to $0$ and $180$ degrees in the probability
density functions of the intermolecular angle; they correspond to aligned molecules. Given the curve in Fig.~\ref{fig:pdfAngle} (a),
two molecules were considered to be aligned if the angle $\theta$ between them satisfies:
\begin{equation}
\theta \leq 34 \text{ degrees} \quad \text{ OR }\quad \theta \geq 149 \text{ degrees}.
\end{equation}
This criterion enables us to distinguish between case (a), where two molecules are perfectly aligned and case (b),
where molecules are not aligned enough, in Fig.~\ref{fig:scheme}.

To quantify how close two aligned molecules should be in order to be declared nearest neighbors
in the same nanoaggregate, we studied the so-called "stacking distance". The stacking distance between
two aligned molecules 1 and 2 is defined in the following way. Molecules 1 and 2 need to be aligned
according to the previous criterion on the alignment angle $\theta$. The definition of the stacking distance
makes then use of the intraplanar distance $d_\text{intra}$, defined as the distance in plane 1 between the center of mass of molecule
1 and the intersection point between plane 1 and the perpendicular line to plane 1
passing through the center of mass of molecule 2 (see Fig.~\ref{fig:scheme} (c) and (d)).
The stacking distance between the aligned molecules 1 and 2 is equal to the distance between the planes where these molecules lie
if the intraplanar distance $d_\text{intra}$ is sufficiently small, otherwise it is equal to the direct
distance between the molecules' centers of mass. The rule for the intraplanar distance $d_\text{intra}$ to be small enough is:
\begin{equation}
d_{\text{intra}} \leq 0.7 \times d_{\text{asph}},
\end{equation}
where $d_{\text{asph}}$ is the size of the body of an asphaltene molecule, which is $d_{\text{asph}} = 13.1$ \AA.
Defining such a stacking distance is useful for two main reasons.
First, studying the distance between the planes instead of the direct distance between the molecules' center of mass is necessary
to find the same distance between two asphaltene molecules aligned on top of each other in a head-to-head conformation and two
asphaltene molecules aligned in a head-to-tail conformation.
These two situations correspond to cases (a) and (c) in Fig.~\ref{fig:scheme}.
As the molecule's center of mass
is not the geometrical center of the molecule, the distance between the centers of mass is different in the case
of a head-to-head and a head-to-tail conformation. However, the distance between the two planes is the same
in both cases and should be considered to declare that, in both cases, the two molecules are nearest neighbors in the
same nanoaggregate.
Nevertheless, considering only the plane distance can lead to vanishing
distances when two molecules happen to be aligned in the same plane but far away from each other.
This situation corresponds to case (d) in Fig~\ref{fig:scheme}.
The threshold value $0.7 \times d_{\text{asph}}$ on the intraplanar distance
is there to rule out this case from the nanoaggregate definition. In that case, the stacking
distance is just the distance between the molecules' centers of mass and is quite large.
The precise value $0.7 \times d_{\text{asph}}$ was chosen because for the smaller value $0.5 \times d_{\text{asph}}$
the number of distances close to zero  was too high and because for the larger value $0.9 \times d_{\text{asph}}$
the asphaltene molecules in a head-to-tail conformation were often computed to be at $9$ \AA $ $ from each other.

\begin{figure}
  \scalebox{0.40}{\includegraphics[angle=-90]{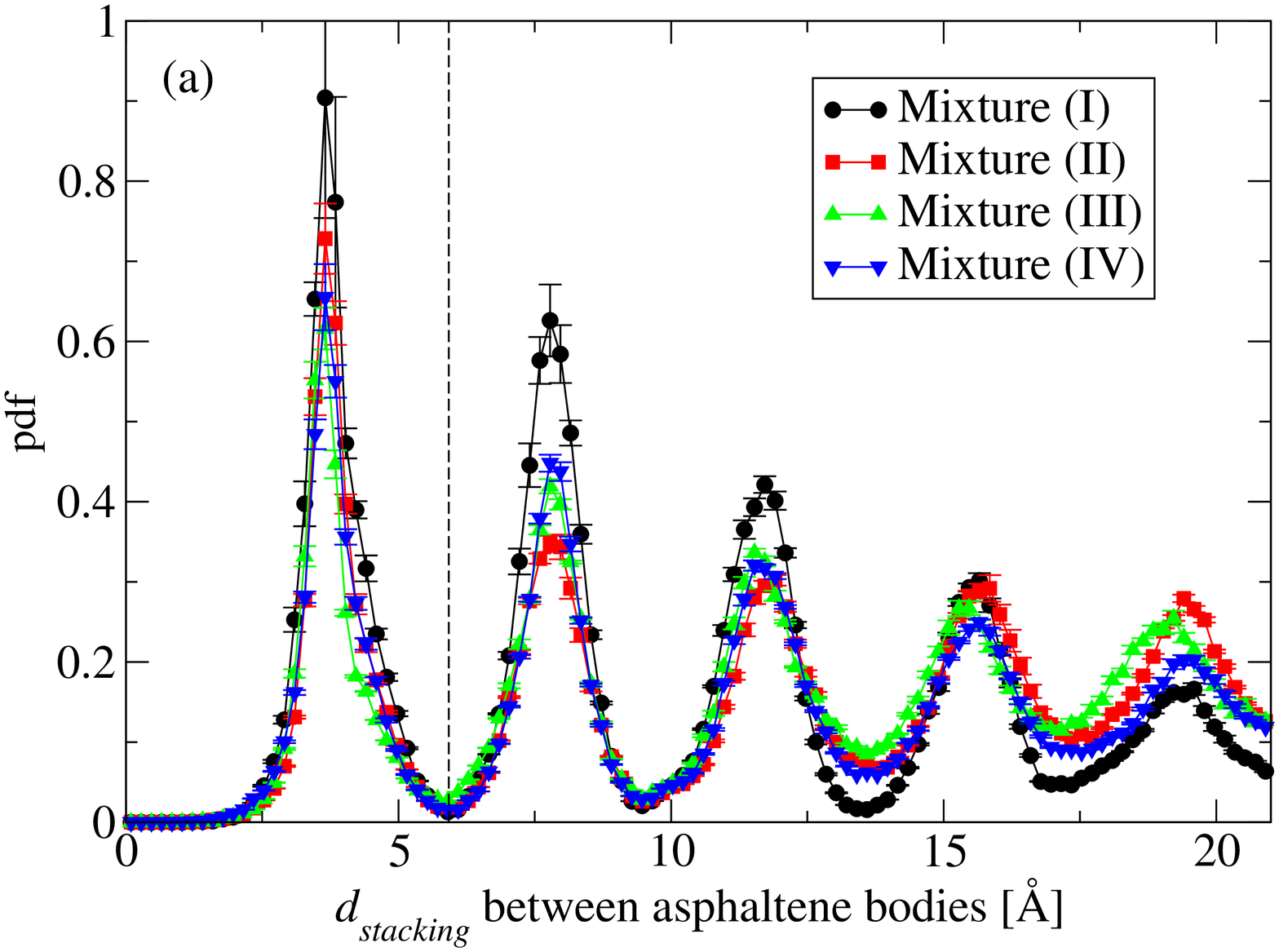}}\\
  \scalebox{0.40}{\includegraphics[angle=-90]{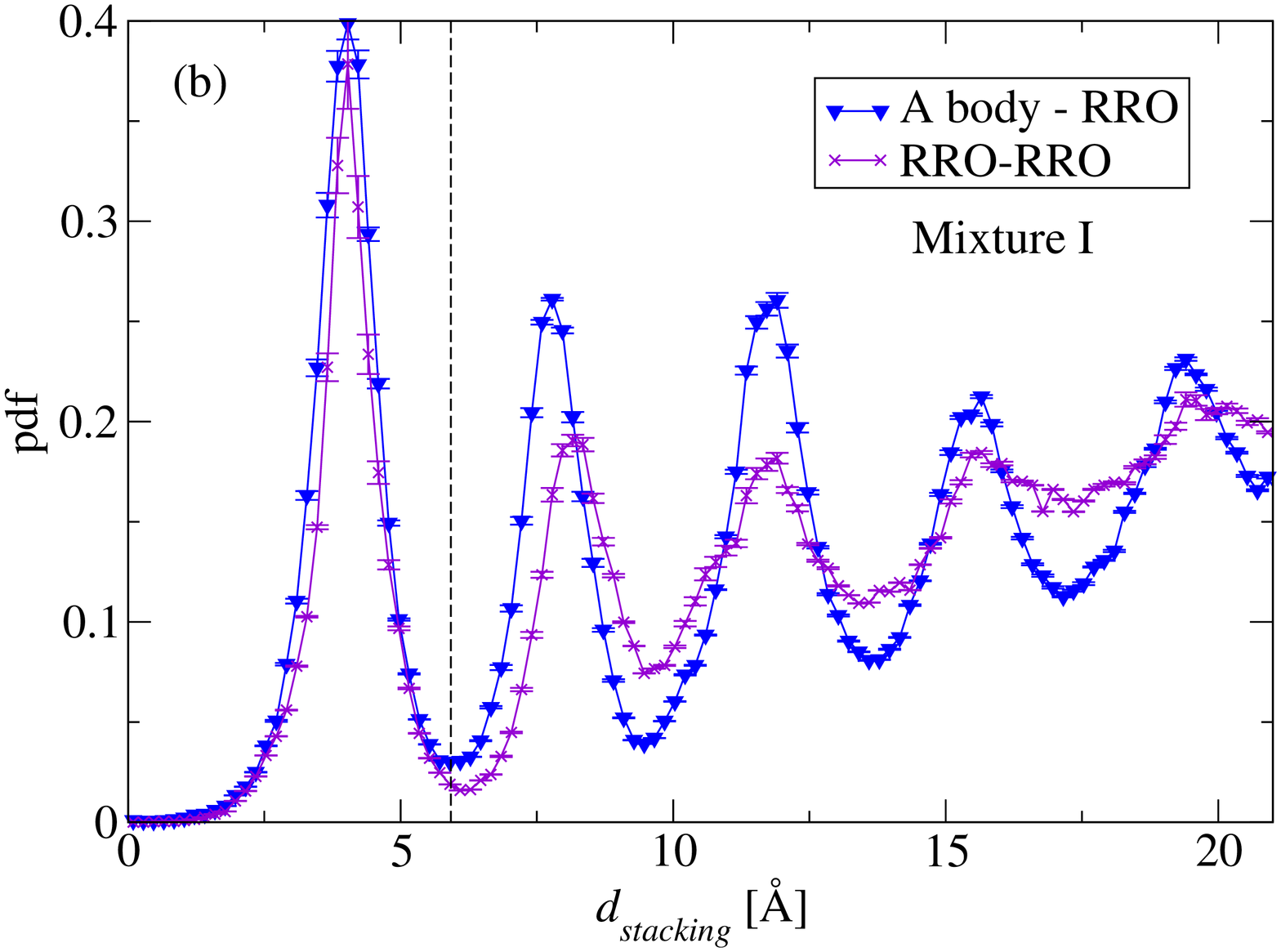}}
  \caption{\label{fig:pdfDistance}
(Color online) (a) Probability density functions (pdfs) of the stacking distance between asphaltene bodies as defined in the text.
(b) Probability density functions (pdfs) of the stacking distance in mixture I
between resin/resinous oil molecules and asphaltene bodies (blue triangle down) and between resin and resinous oil
(purple cross).
The dashed line in both figure represents the last threshold value chosen (6.0 \text{\AA}).
}

\end{figure}

Figure~\ref{fig:pdfDistance} represents the probability density function of the stacking distance between two aligned asphaltene bodies
defined from the two previous threshold values in the four
different mixtures. The extended stacked structure is visible in this figure whereas it was not so clear in
the radial distribution function of the distance between molecules centers of mass shown in Fig.~\ref{fig:rdf} (a). 
The regular and well-defined peaks of Fig~\ref{fig:pdfDistance} justifies \emph{a posteriori} the two threshold values defined earlier
on the alignment angle and the intraplanar distance.
One last threshold value can be inferred from this figure. Two aligned molecules are said to be close enough if
the stacking distance $d_{\text{stacking}}$ between them is smaller than:
\begin{equation}
d_{\text{stacking}} \leq 6.0 \text{\AA}.
\end{equation}
In order to check the choice of the two last threshold values
for pairs of molecules different from two asphaltene bodies, we plotted the probability density functions of the stacking distance
between other types of molecule. As an example, Fig.~\ref{fig:pdfDistance} (b) represents the pdf
of the stacking distance in mixture I between resin and resinous oil molecules on one hand
and a resin/resinous oil molecule and an asphaltene body on the other hand. The stacked structure is less obvious than
for the asphaltene body distance, but the first peak is still very well-defined.
The minimum after the first peak is very close to the minimum observed in the pdf of the stacking distance between
two asphaltene bodies. This justifies the use of the same
threshold values for the resin/resinous oil and resin/resinous oil asphaltene
pairs. The same conclusion can be drawn for all the other types of pairs (not shown).

\begin{figure}
\includegraphics[angle=-90, scale=0.4]{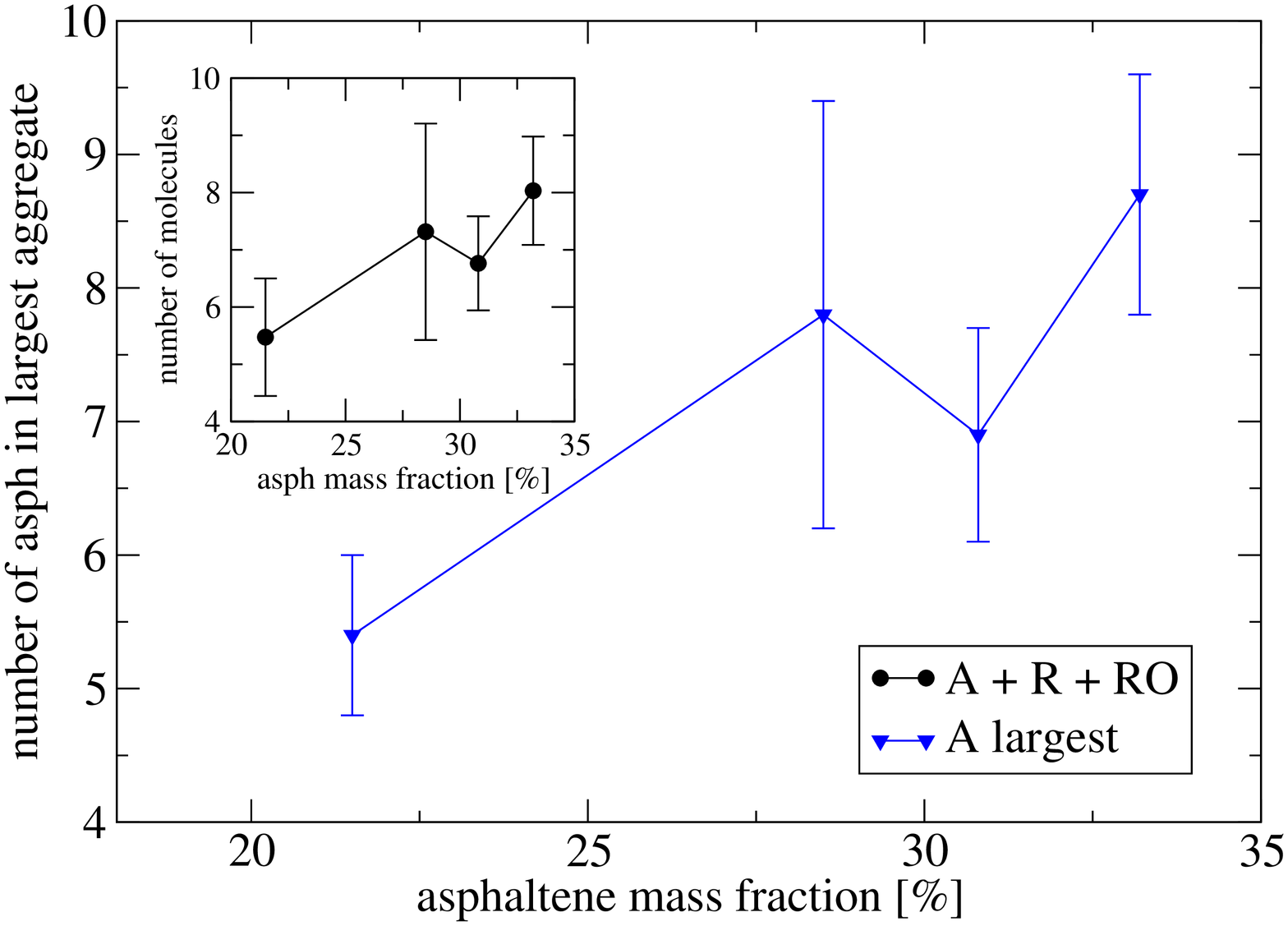}
\caption{
Variation of the average number of asphaltene molecules
in the largest nanoaggregate versus the asphaltene mass fraction in the different mixtures. 
Inset: Variation of the average number of molecules (asphaltene, resin, and resinous oil)
in any aggregate versus the asphaltene mass fraction in the different mixtures.
}
\label{fig:aggregateSize}
\end{figure}

With this precise molecular definition of an asphaltene nanoaggregate, the average number of asphaltene molecules in the largest
aggregate can be computed for the different mixtures. The smallest nanoaggregate considered contains
at least two asphaltene molecules, which can be indirectly linked by one or several resin or resinous oil molecules.
The number of asphaltene molecules in the largest nanoaggregate is determined regularly in each simulation for each mixture.
The average number of asphaltene molecules in the largest nanoaggregate is plotted versus
the mass fraction of asphaltene molecules in the different mixtures in Fig.~\ref{fig:aggregateSize}.
The largest nanoaggregate is found to contain between 5 and 10 asphaltene molecules, which is
an agreement with the experimental literature reporting less than ten asphaltene molecules 
on average in a nanoaggregate~[\onlinecite{mullins2011}].
Figure~\ref{fig:aggregateSize} also shows that the average number of asphaltene molecules
inside the largest nanoaggregate increases going from mixture I to IV, i.e. as the bitumen ages.
It is not surprising that the number of asphaltene molecules inside a nanoaggregate increases as bitumen ages,
because the number of asphaltene molecules in the whole mixture increases. But the study carried out here
quantifies properly this effect.
Moreover, we have checked that the average number of aromatic molecules (asphaltene, resin and resinous oil)
inside any nanoaggregate increases as bitumen ages. The result is displayed in the inset of Fig.~\ref{fig:aggregateSize}.
Even if the error bars are quite large, there is a definite increase in the aggregate size from mixture
I to IV. This is even more surprising given that the total number of aromatic molecules decreases from mixture I to IV.
It means that the higher the concentration of asphaltene molecules is in the whole mixture, the easier it is
for them to recruit themselves and other aromatic molecules to a nanoaggregate.
One can now argue that the slowdown in the stress autocorrelation
function dynamics observed in section~\ref{sec:stress} is correlated to an increase in the nanoaggregates' size
and an increase in their asphaltene content.

The question whether the layered structure composed of asphaltene, resin and resinous oil molecules is a nanoaggregate miscible in docosane or
a separate phase cannot be directly addressed by molecular dynamic simulations, because the systems that can be studied numerically are too small.
Consequently, the results of molecular dynamic simulations presented here can be influenced by finite size effect. However
the general trends are probably the same in larger systems as they agree with experimental results on the nanoaggregate size~[\onlinecite{mullins2011}].

\begin{figure}
\includegraphics[angle=-90, scale=0.4]{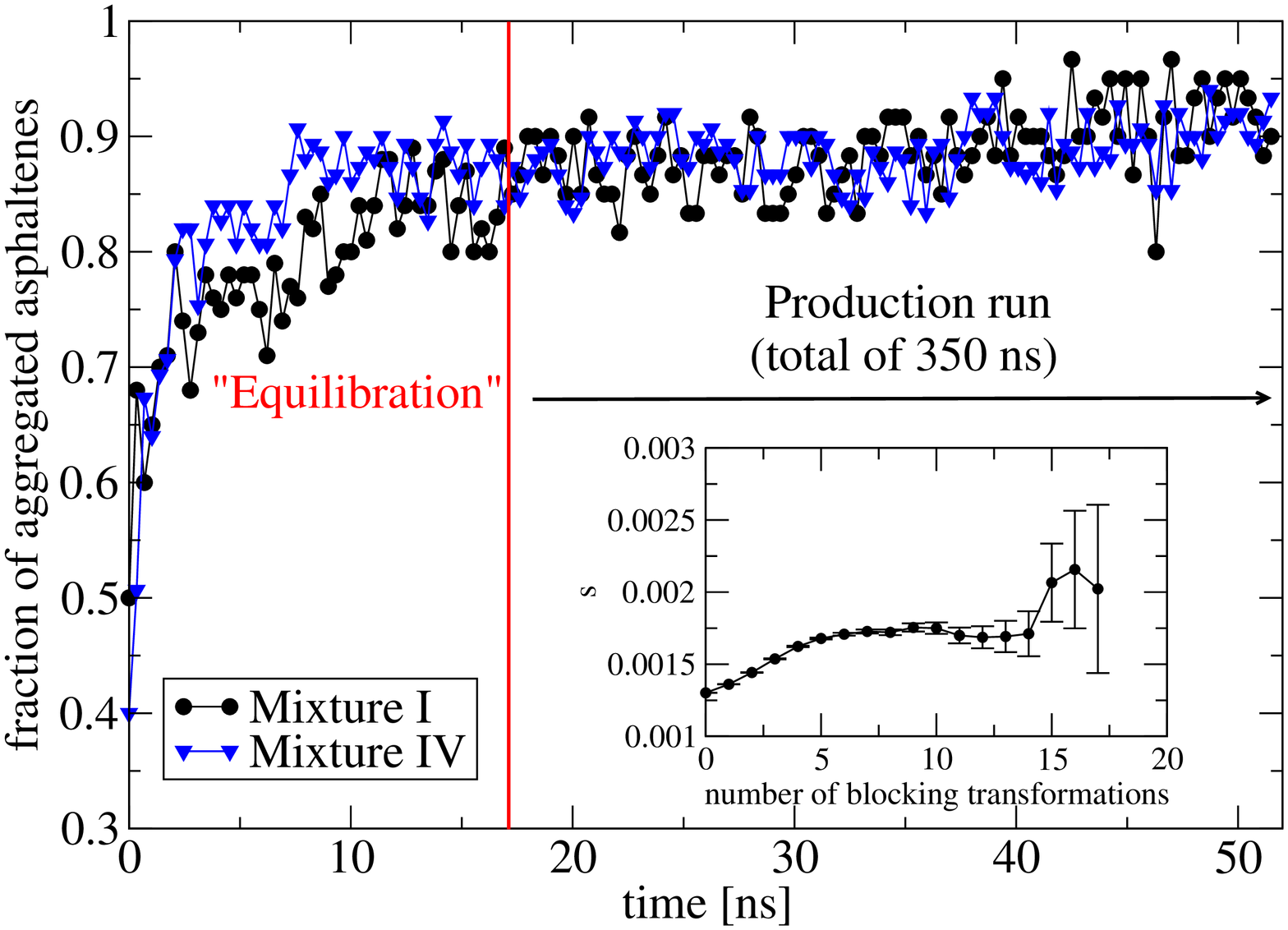}
\caption{
(Color on line) Time evolution of the fraction of aggregated asphaltene molecules for
mixture I (black circle) and IV (blue triangle down).
Inset: results of the blocking method applied to pressure for one production run of mixture I.
}
\label{fig:nbAggregatedAsph}
\end{figure}

The proper definition of a nanoaggregate allows us to justify \emph{a posteriori} the time span used
to "equilibrate" the system. Figure~\ref{fig:nbAggregatedAsph} shows the fraction of aggregated asphaltene molecules
among all asphaltene molecules versus time for mixtures I and IV. During the "equilibration run"
the number of aggregated asphaltene molecules increases until it reaches a stationary value. 
The number of aggregated asphaltene molecules increases because the initial configuration is generated
from a very dilute solution and then compressed.
The production
runs start when the stationary value for the number of aggregated asphaltene molecules
is reached. 
Moreover, to check that the production runs span the phase space properly, we used the standard
"blocking method"~[\onlinecite{flyvberg89}]. 
We applied this method to the pressure and computed the standard deviation $s$ of the average
pressure versus the number of blocking transformations. The results
are displayed for one production run of mixture I in the inset of Fig.~\ref{fig:nbAggregatedAsph}.
A plateau is clearly visible after a few blocking transformations have been applied, indicating
that the phase space is spanned correctly during a production run.
We assumed, in agreement with the results from the blocking method, that the state
describe by the molecular dynamics simulations is a state of local equilibrium.
It is associated with a given nanoaggregate configuration.
We assumed that the laws of statistical mechanics applied to this local equilibrium state.
To span different nanoaggregate configurations, we simulated eight to ten systems with
different initial conditions and always considered quantities averaged over these different simulations.

To summarize, in this section the nanoaggregates were identified in all mixtures. They are not only composed of asphaltene molecules:
resin and resinous oil molecules align to the asphaltene structure. A precise definition for the nanoaggregate
structure was arrived at and it was shown that the slowdown of the stress autocorrelation function observed
as bitumen ages is correlated to an increase in the nanoaggregate size. To deeper investigate
the correlation between the evolution of mechanical properties and asphaltene aggregation as bitumen ages,
the next section is devoted to the study of the nanoaggregates' dynamics. 

\section{Rotational dynamics}
\label{sec:rot}

It is known from section~\ref{sec:stress} that as bitumen ages, the stress autocorrelation function dynamics is slowed down on average.
This is correlated to nanoaggregates being larger and containing more asphaltene molecules.
To show that it is also linked to a slowdown in the nanoaggregates' dynamics, 
we investigated the evolution of the rotational dynamics of the aromatic molecules in a nanoaggregate.

\begin{figure}
\includegraphics[angle=-90, scale=0.4]{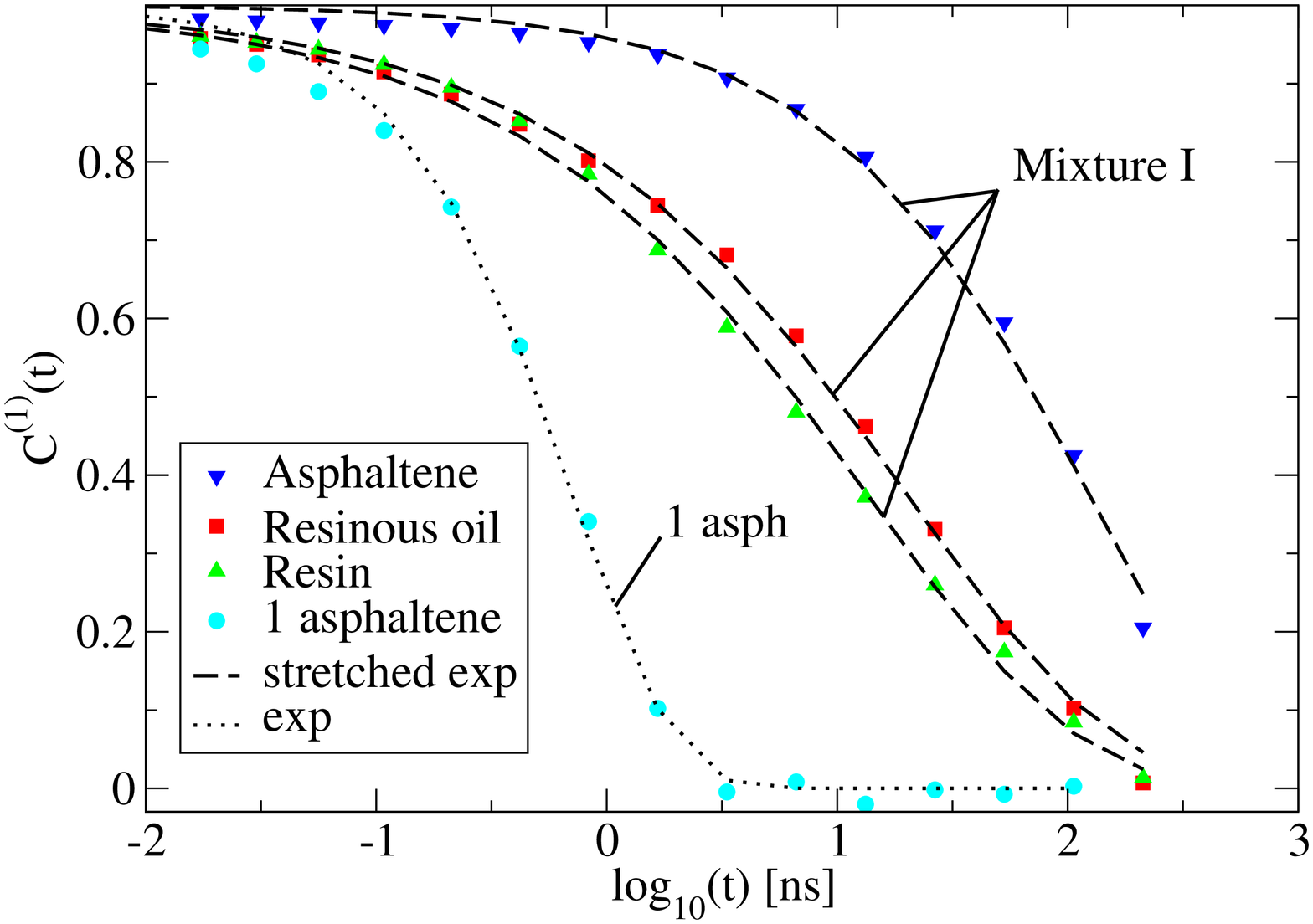}
\caption{
(Color online). Time evolution of the logarithmic average of the first-order rotational correlation function $C^{(1)} (t)$
in mixture I and in a mixture containing one asphaltene molecule and $82$ docosane molecules.
Blue triangles down correspond to asphaltene in mixture I,
red squares to resinous oil in the same mixture, and green triangles up to resin in the same mixture.
The three dashed black lines are stretched exponential fit to the curves.
Light blue circles correspond to asphaltene in a mixture containing one asphaltene molecule in docosane.
The dotted black line is the best fit of the form $C^{(1)} (t) = \exp(-2D_rt)$, with $D_r = 0.83$ ns$^{-1}$, the
rotational diffusion coefficient.
}
\label{fig:rotDyn}
\end{figure}

The normal unit vectors to the asphaltene bodies and resin and resinous oil molecules defined in section~\ref{sec:structure}
can be used to investigate the rotational dynamics of each molecule.
In other words, they can be used to determine how long the initial orientation is preserved.
To do so, the first-order rotational correlation function $C^{(1)} (t)$ is defined as~[\onlinecite{hansen}]:
\begin{equation}
C^{(1)}(t) = \langle \hat{\mathbf{n}}(0) \cdot \hat{\mathbf{n}}(t)\rangle,
\end{equation}
where $\hat{\mathbf{n}}(t)$ is the unit vector normal to the plane aromatic structure
of the molecule of interest at time $t$ and where $\langle\cdot\rangle$ denotes an ensemble average.
In our simulations, the average is done over the molecules of the same type, over time, and over the eight to ten independent
simulations performed for the same mixture.
The rotational correlation function $C^{(1)} (t)$ is plotted in Fig.~\ref{fig:rotDyn} for
asphaltene, resin, and resinous oil molecules in mixture I. 
It can be seen from this figure that the rotational correlation function of the different molecule types
can be fitted to a stretched exponential of the form $C^{(1)}(t) = \exp(-(t/\tau)^{\beta})$, where
$\tau$ can be interpreted as a relaxation time and $\beta$ is a number smaller than $1$. The mean values
of the power $\beta$ in the stretched exponential are $\beta = 0.65 \pm 0.10$ for the asphaltene type,
$\beta = 0.52 \pm 0.08$ for the resinous oil type, and $\beta = 0.45 \pm 0.07$ for the resin type. The average is over
ten independent simulations.
These values are significantly smaller than unity, whereas the rotational correlation function
of a mixture with one asphaltene molecule in docosane can be very well fitted by a simple exponential ($\beta$ = 1)
for sufficiently large time,
as can also be seen in Fig.~\ref{fig:rotDyn}.
This proves that it is the presence of several asphaltene molecules that induces values of the power $\beta$
smaller than unity and not the internal structure of one asphaltene molecule.
 
\begin{figure}
\includegraphics[angle=-90, scale=0.4]{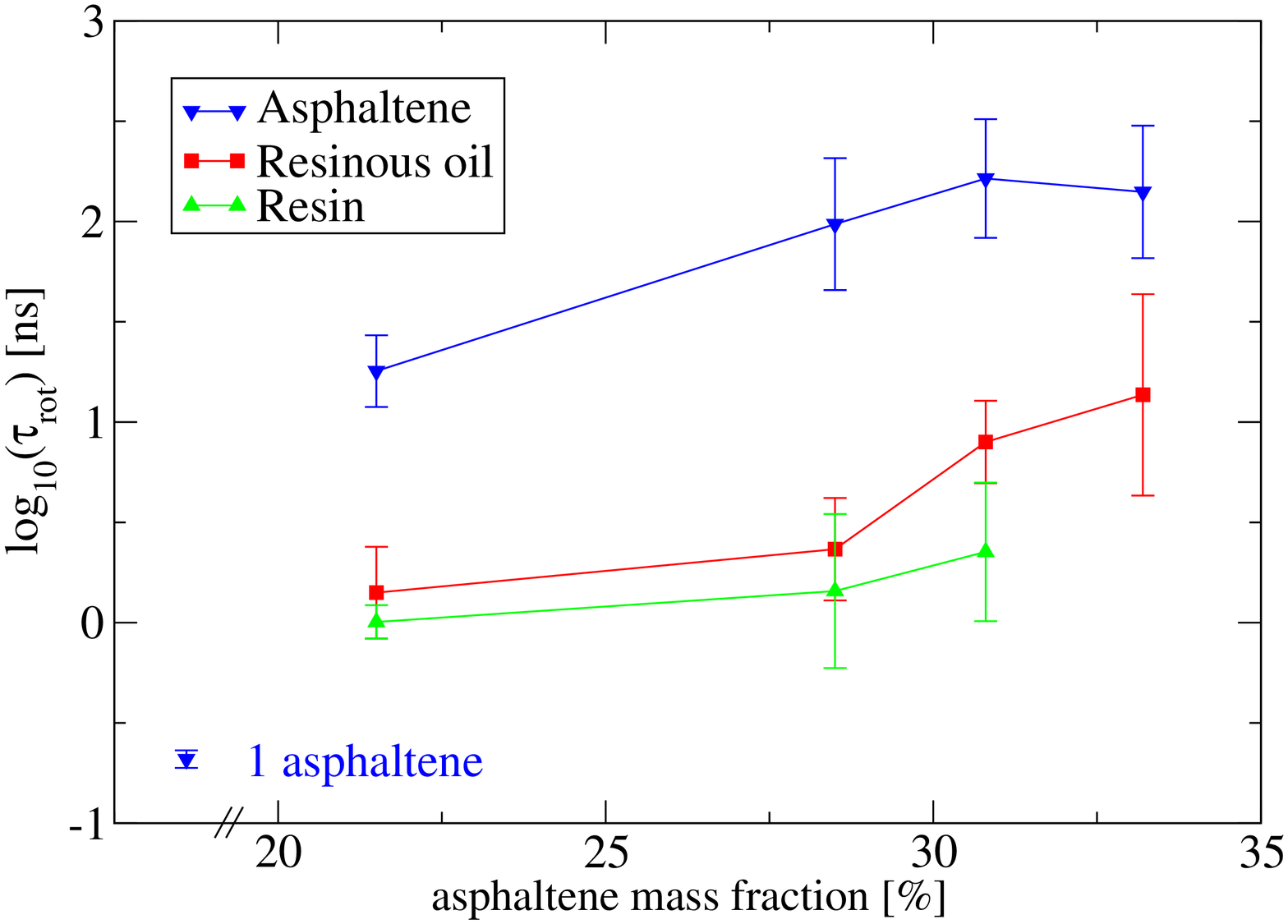}
\caption{
(Color online). Variation of the rotational characteristic time $\tau_{\text{rot}}$, defined as $C^{(1)} (\tau_{\text{rot}}) = 3/4$,
versus the number of asphaltene molecules in the mixture.
The rotational characteristic time indicated for one asphaltene corresponds to a mixture with one asphaltene molecule
and $82$ docosane molecules, as in Fig.~\ref{fig:rotDyn}.
Error bars correspond to the standard deviations computed over the 8 or more independent simulations performed in each case.
}
\label{fig:rotTime}
\end{figure}

For mixtures II to IV the rotational correlation functions $C^{(1)} (t)$ can also be computed for the
different molecule types, but
stretched exponential fits were found to be less convincing. To circumvent this problem, we
chose to define the characteristic rotational time $\tau_{\text{rot}}$ as the time necessary
for the rotational correlation function $C^{(1)} (t)$ to reach the value $3/4$.
The variation of the rotational time $\tau_{\text{rot}}$ versus the asphaltene mass fraction in the mixture
is shown in Fig.~\ref{fig:rotTime}, for all molecule types.
This figure shows that the asphaltene characteristic rotational time is the largest. This is expected as they are the largest molecules.
Asphaltene molecules are the slowest and impose their dynamics on the whole mixture.
Resin and resinous oil molecules have smaller characteristic rotational times, quite close to each other.
Interestingly, Fig.~\ref{fig:rotTime} also shows that all characteristic rotational times increase on average
as bitumen ages. 
It means that, as bitumen ages, the rotational dynamics is slowed down.
There is consequently a correlation between the slowdown in the stress autocorrelation function observed in section~\ref{sec:stress} and
the slowdown in the rotational dynamics, as bitumen ages. 

The last sections indicated that asphaltene molecules impose their structure and dynamics on the whole mixture.
The role of resin and resinous oil molecules, which are also part of the nanoaggregates, is however not completely clear.
The next section focuses on the role of these two molecules and studies the translational diffusivity of each molecule
type as bitumen ages.

\section{Translational diffusion}
\label{sec:diff}
The previous sections~\ref{sec:stress},~\ref{sec:structure}, and ~\ref{sec:rot} showed that as bitumen ages, the stress autocorrelation
function decays more slowly while nanoaggregates become larger and the rotational dynamics becomes slower.
The role of asphaltene molecules seems critical in this process, because they are the largest molecules
and seem to impose their structure and dynamics on the whole mixture.
But the role of resin and resinous oil molecules in the nanoaggregate formation is not yet clear.
To specify the role of each molecule type,
this section studies the diffusivity of these different molecule types.

\begin{figure}
  \scalebox{0.25}{\includegraphics[angle=-90]{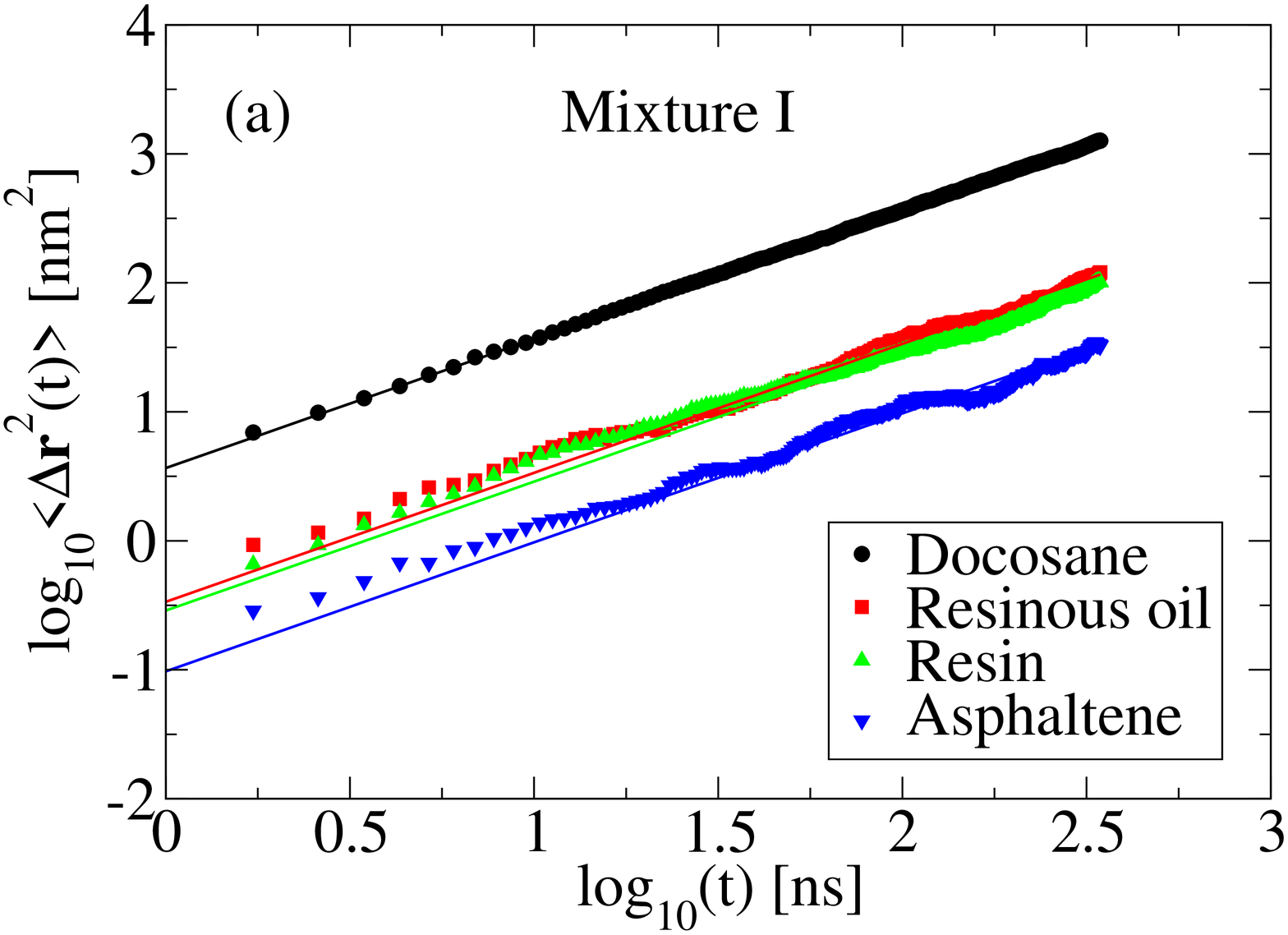}}
  \scalebox{0.25}{\includegraphics[angle=-90]{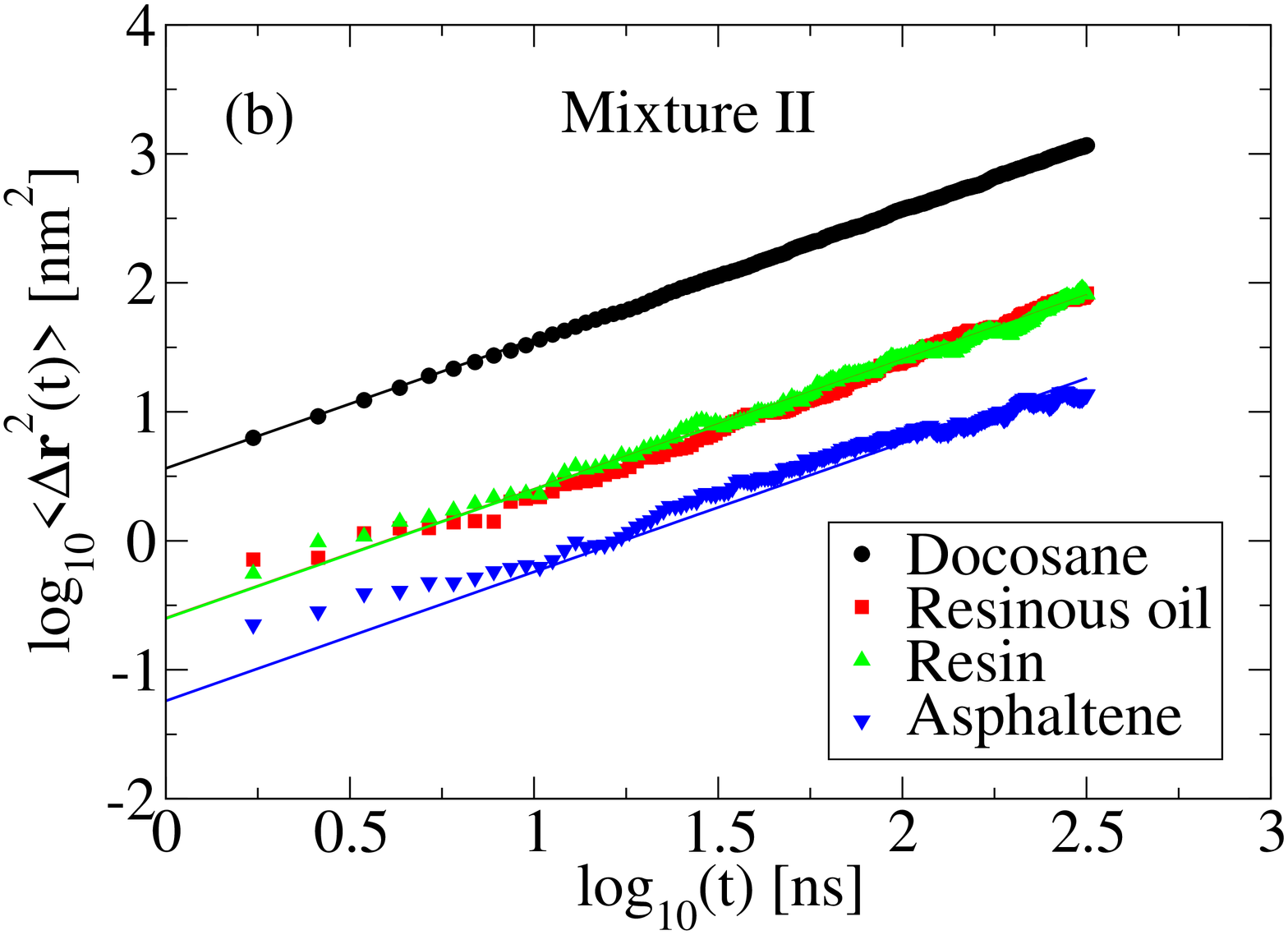}}\\
  \scalebox{0.25}{\includegraphics[angle=-90]{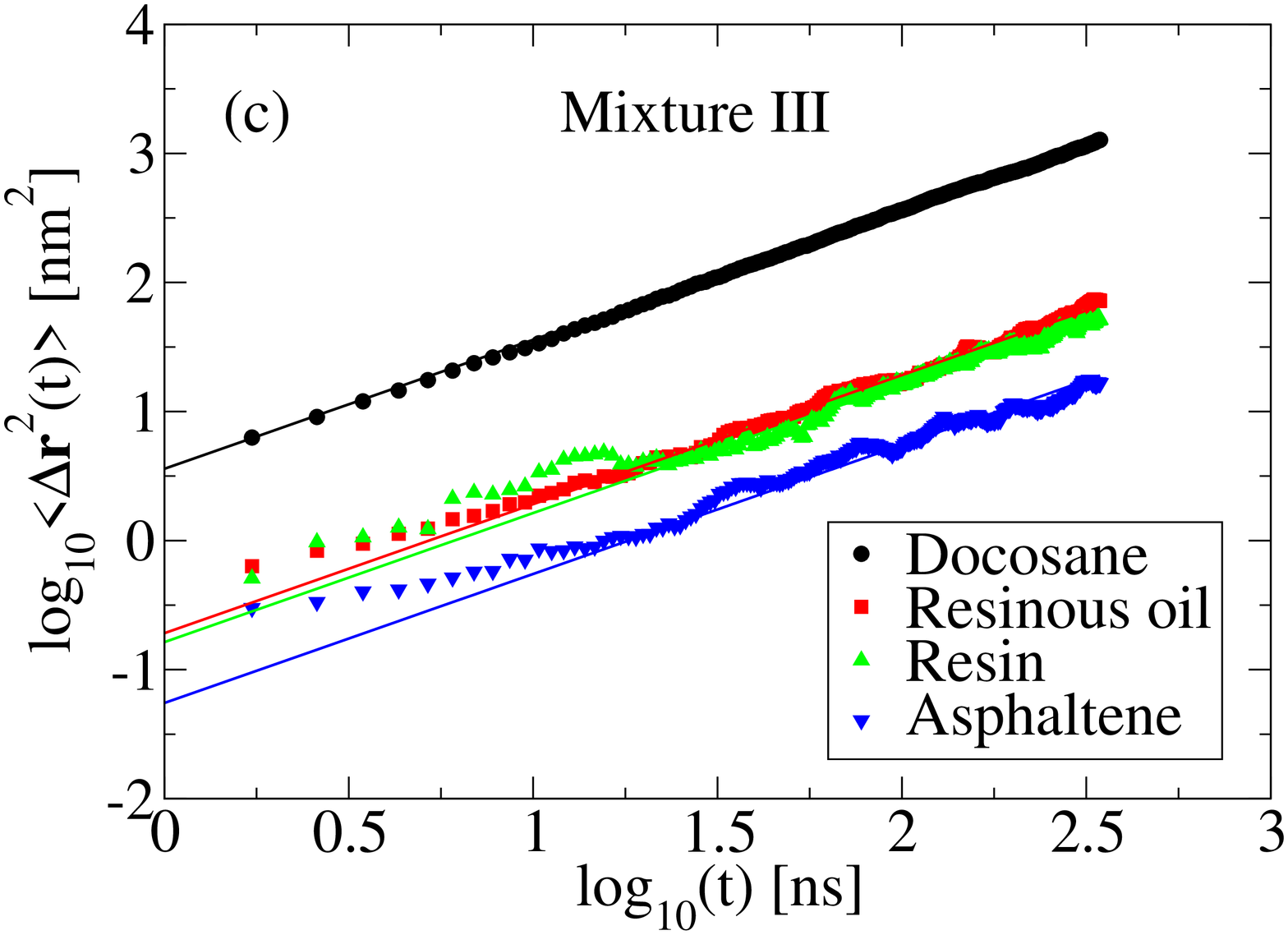}}
  \scalebox{0.25}{\includegraphics[angle=-90]{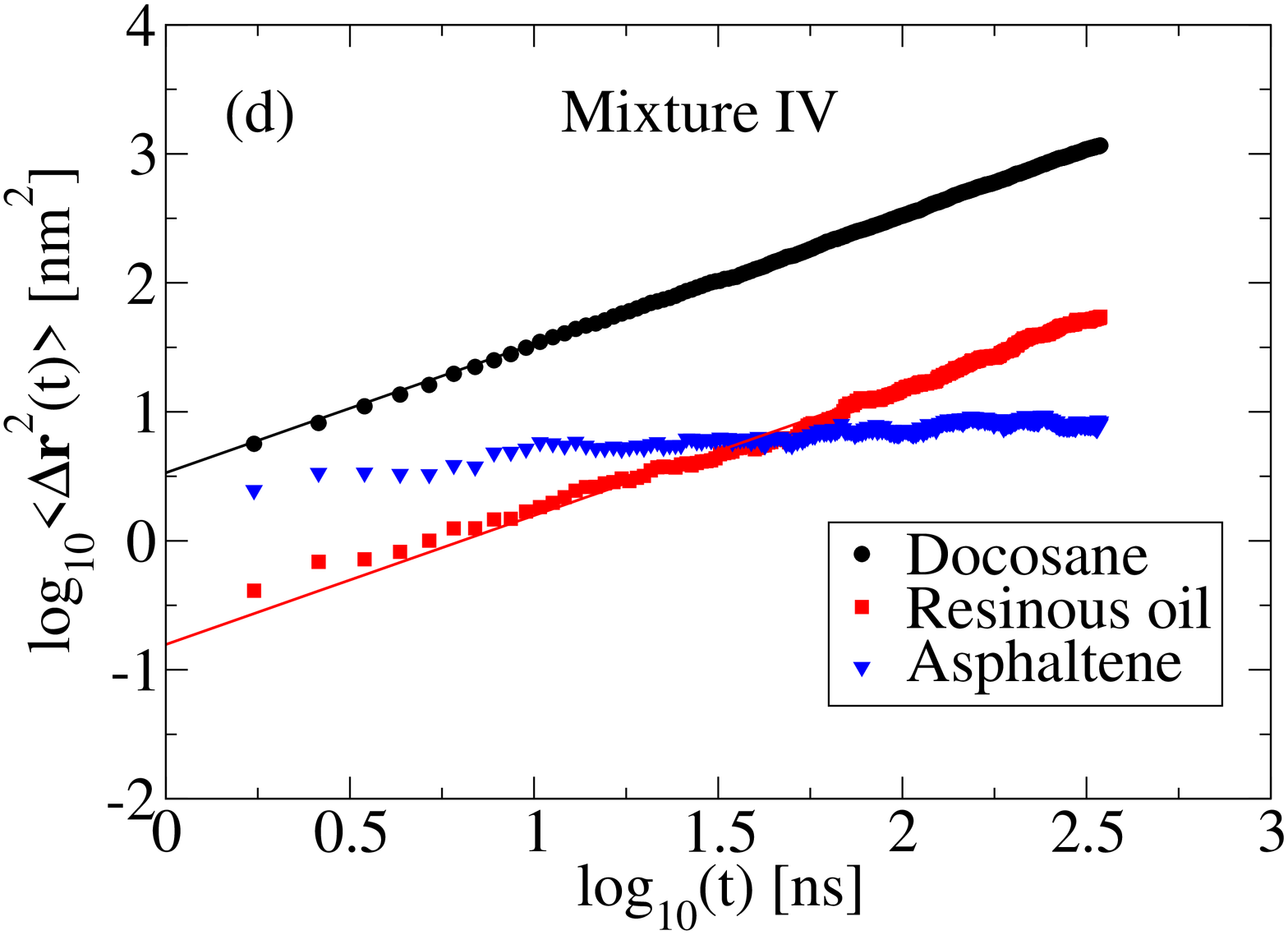}}
  \caption{\label{fig:MSD}
    (Color online). Time evolution of the mean-squared displacement of each molecule in all four mixtures.
The curves are averaged over the independent simulations performed for each mixture: 10 for mixture I,
8 for mixture II, 9 for mixture III, and 10 for mixture IV.
}

\end{figure}

The mean-squared displacement $\langle \Delta \mathbf{r}^2\rangle$ is defined for one molecule type from
the center-of-mass positions as in Ref.~[\onlinecite{us}].
The curves displayed in Fig.~\ref{fig:MSD} show the time evolution of the average of the mean-squared displacement
over the eight or more independent simulations performed for each mixture.

In the first three mixtures, all molecule types exhibit a diffusive behavior.
In that case a diffusion coefficient $D_{\alpha}$ can be defined for each molecule type.
In mixture IV, which does not contain any resin molecules, the docosane and resinous oil molecules still exhibit a diffusive behavior, as shown
in Fig.~\ref{fig:MSD}, but the asphaltene molecules do not enter the diffusive regime, even after $0.35$ $\mu$s. The asphaltene
molecules appear stuck in their initial position for the time spans accessible to molecular dynamics. 

\begin{figure}
\includegraphics[scale=0.4, angle=-90]{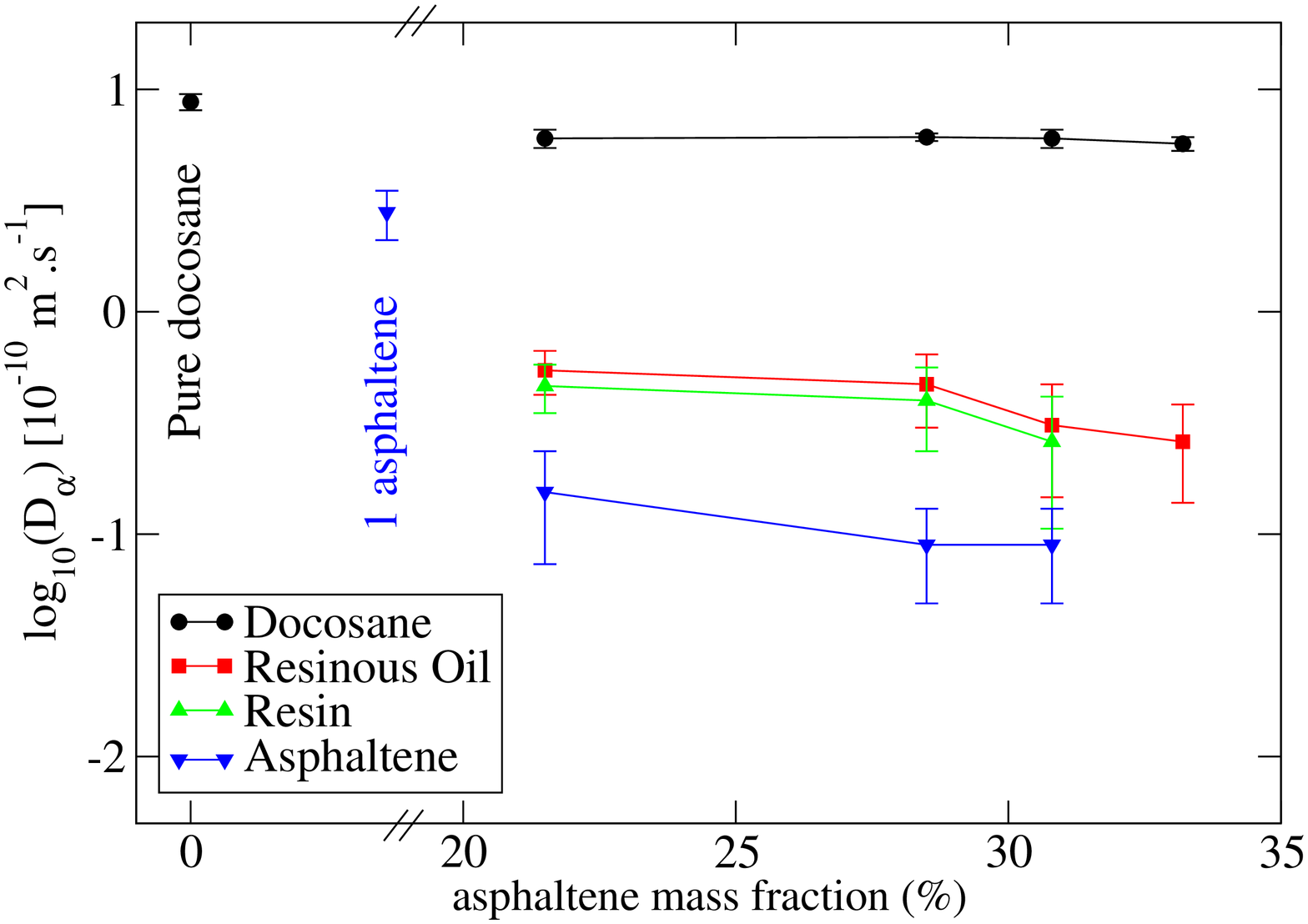}
\caption{
(Color online). Variation of the diffusion coefficient $D_{\alpha}$ of each molecule type versus the asphaltene mass fraction
in the different mixtures.
The same symbols are used as in Fig.~\ref{fig:MSD}.
Error bars correspond to the standard deviations computed over the eight or more independent simulations performed in each case.
}
\label{fig:diffCoeff}
\end{figure}

Figure~\ref{fig:diffCoeff} plots the diffusion coefficient of each molecule type versus the asphaltene mass fraction in the mixture
and reveals the role of each molecule type. 
Firstly, different molecule types have different diffusion coefficients. As was already shown in
our previous work~[\onlinecite{us}]: $D_{\text{D}} > D_{\text{RO}} \simeq D_{\text{R}} > D_{\text{A}}$.
Secondly, except for the docosane diffusion coefficient, which is roughly constant, all diffusion
coefficients decrease with an increase in the asphaltene content. The aging reaction consequently
leads to a slowdown in the translational dynamics. It is not surprising as it is also related
to a slowdown in the rotational dynamics and the stress autocorrelation function dynamics, as could be seen
from sections~\ref{sec:stress} and~\ref{sec:rot}.

The fact that docosane molecules diffuse the fastest is expected since they are the lightest and
are not part of the nanoaggregate structure.
The diffusion coefficient of pure docosane is only a bit larger than the diffusion coefficient of this same molecule in a bitumen mixture
as shown in Fig.~\ref{fig:diffCoeff}.
This fact can enable us to design coarse-grained simulations where docosane molecules are only implicitly included as a solvent
damping the other molecules motion. This kind of simulations will be the focus of a later work.

The fact that the asphaltene diffusion coefficient is the smallest agrees with their
characteristic rotational time being the highest. It is expected since they are the largest molecules in the bitumen mixture.
The effect of the asphaltene aggregation on the asphaltene diffusion coefficient is clearly visible when one
compares the asphaltene diffusion coefficient in a mixture with a single asphaltene molecule in pure docosane
and the same diffusion coefficient in a bitumen mixture. When no aggregation takes place, the asphaltene
diffusion coefficient is one order of magnitude higher than when several asphaltene molecules are present and aggregated together.
This is another indication of the importance of asphaltene aggregation in bitumen dynamics.

Resin and resinous oil molecules have close diffusion coefficients, which is consistent with their
rotational relaxation time being close.
However, there is a difference between the rotational and translational dynamics of resinous oil molecules
as bitumen ages. In section~\ref{sec:rot}, in Fig.~\ref{fig:rotTime}, it was shown that the rotational time of
resinous oil molecules is increased by a factor 10 going from mixture I to IV, whereas the diffusion
coefficient of resinous oil molecules is only decreased by a factor two from mixture I to IV.
The same is also true for resin and asphaltene molecules.
This decoupling between the translational and rotational dynamics can be interpreted as the consequence of spatial
heterogeneities~\cite{cavagna} growing with age, consistent with the nanoaggregates being larger as bitumen ages.
The extreme case would be a complete decoupling between regions with free molecules, which contribute
a lot to the diffusive behavior but do not relax fully the rotational dynamics,
and regions with large nanoaggregates, where molecules
do not diffuse and rotate very slowly.

\begin{figure}
\includegraphics[scale=0.4, angle=-90]{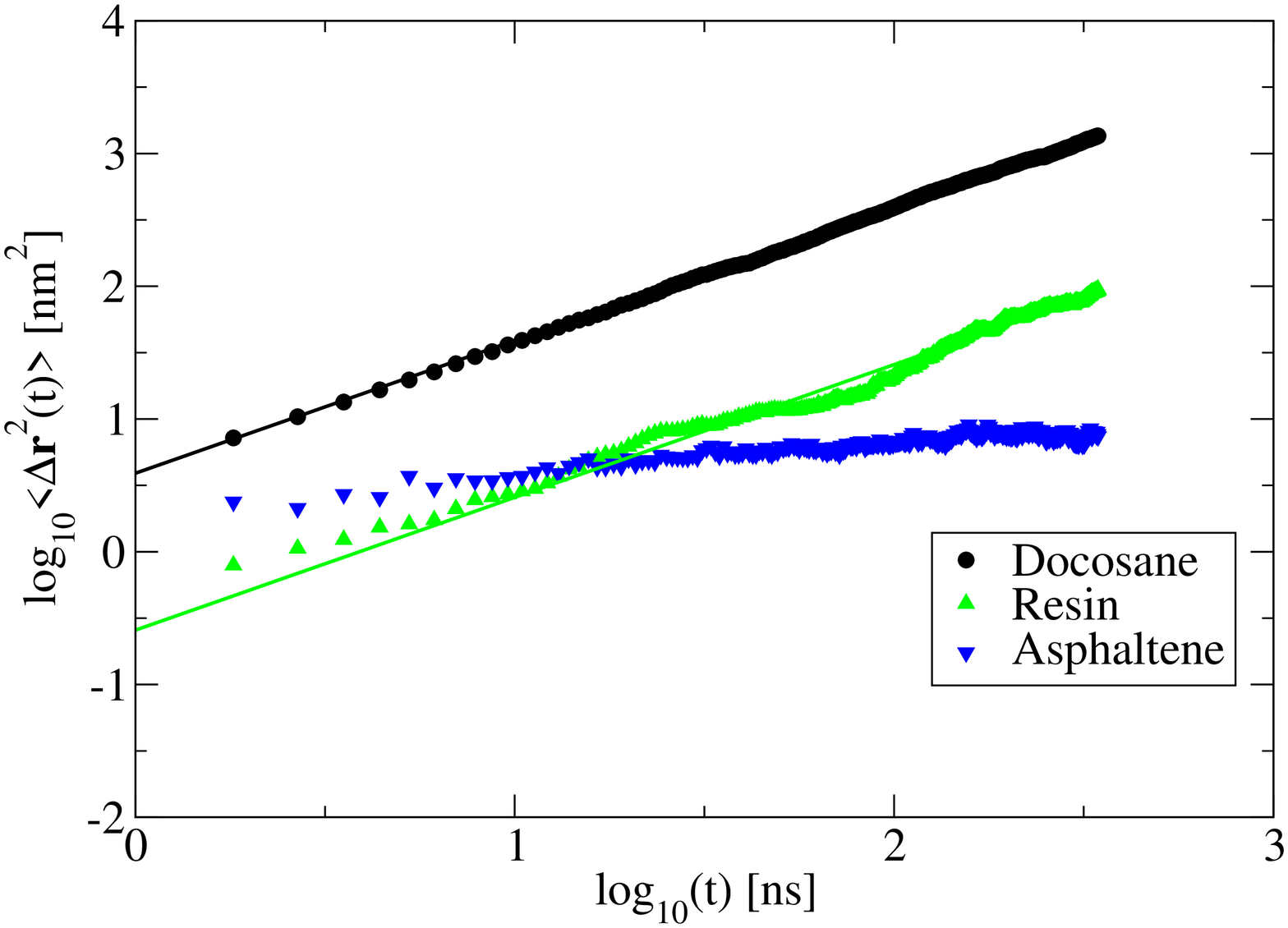}
\caption{
(Color online). Time evolution of the mean-squared displacement of each molecule in a mixture containing
82 docosane molecules, 10 resin molecules and 10 asphaltene molecules (to be compared with Fig.~\ref{fig:MSD} (d)).
The curves are averaged over the ten independent simulations performed for this mixture.
}
\label{fig:msdNoresOil}
\end{figure}

In mixture IV where no resin molecules are present, the asphaltene molecules are not able to 
reach a diffusive behavior in the time span accessible to molecular dynamics, whereas they were so with only a couple of resin molecules
in mixture III. 
To test whether it is a property of resin molecules alone, we studied the mean-squared displacement of all molecules
in a mixture containing 82 docosane molecules, 10 resin molecules and 10 asphaltene molecules, but no resinous oil molecule.
The mean-squared displacement of docosane, resin, and asphaltene molecules is represented versus time
in Fig.~\ref{fig:msdNoresOil}. This figure shows that with resin molecules but without resinous oil molecules,
asphaltene molecules do not reach a diffusive behavior in the time span accessible to molecular dynamics.
The other molecule types, on the contrary, have a diffusive behavior.
Resin and resinous oil molecules seem to play similar roles in our simulations.
A sufficient amount of them is needed to induce
asphaltene molecules to have a diffusive behavior.
It means that these molecules either enable single asphaltene molecules to detach from the aggregates from time to time
or induce smaller aggregates to be formed. 

$ $\\

Collecting all the results, we propose the following simple scenario for the role of each molecule type in our model bitumen.
Asphaltene molecules are the largest and have an extended aromatic flat structure. They align and create basic
nanoaggregates. Resin and resinous oil molecules also have a flat aromatic structure, and they can align with the asphaltene
molecules; they are also members of the nanoaggregates. 
However, resin and resinous oil molecules appear to prevent the asphaltene molecules from being immobile. 
This fact may be due to
resin and resinous oil molecules being smaller than the asphaltene molecules and weakening the nanoaggregate structure.
Moreover,
the resin molecules that were chosen in this model have long side chains, that 
cannot align with the aromatic planes and can disturb the layered aggregate structure. 
Resin and resinous oil molecules are believed to be able to break off an asphaltene molecule from a nanoaggregate, because they can
insert between two asphaltene molecules in an aggregate thanks to their flat aromatic structure and also disturb the aggregate layers
because of their smaller size. 
Finally, docosane molecules are not members of the nanoaggregates and their diffusion coefficient does not depend on the precise
chemical composition of bitumen.

With this scenario in mind, one can explain qualitatively what happens in the bitumen mixture as the aging reaction
"2 resin $\rightarrow$ 1 asphaltene" takes place. As resin molecules are replaced
by asphaltene molecules, the nanoaggregates become larger
because resin molecules do not disturb the aggregate structure any more. The translational and rotational dynamics
of all aromatic molecules is consequently slowed down. This also affects the stress autocorrelation function dynamics,
which is slowed down, resulting in an increase in viscosity.

\section{Summary}
\label{sec:conclu}

We have shown that as the number of asphaltene molecules increases and the number of resin molecules decreases,
that is to say as bitumen ages, the decay of the stress autocorrelation function
gets slower. This is consistent with the experimental literature on aged bitumen reporting an increase in
viscosity~[\onlinecite{shrp368, herrington, lu, ec140}]. The aging reaction has an important and
significant influence on bitumen rheological properties in our model.
We have also shown that the stress autocorrelation dynamics slowdown is correlated to the formation of
larger nanoaggregates and to a slowdown in the rotational dynamics.
Finally, we described qualitatively the role of each molecule type in our model bitumen. Asphaltene molecules tend to aggregate
forming a stacked structure. 
Docosane molecules are not members of this structure.
Resin and resinous oil molecules follow the structure and align with the asphaltene molecules.
In addition, they are able to break the structure and detach some
asphaltene molecules from it.
As bitumen ages, the aggregation process is enhanced, because resin molecules disappear.
This mechanism explains why asphaltene nanoaggregates get larger as bitumen ages.
It results in a dynamic slowdown and in a viscosity increase.

Knowing this scenario enables us to propose a simple way to rejuvenate bitumen: adding small aromatic molecules
such as resin and resinous oil molecules to the mixture. These additives are hopefully cheaper than
polymer additives as they are naturally present in bitumen.

The interaction between resin and asphaltene molecules is a longstanding issue in the bitumen literature~[\onlinecite{mullins2010}].
Resin molecules were first thought to act as surfactants coating clusters of asphaltene nanoaggregates. This picture
was disregarded because experimental evidence showed that only a small fraction of resin molecules
are associated to asphaltene nanoaggregates~[\onlinecite{indo}]. We argue here that resin molecules may interact with asphaltene nanoaggregates
in such a way that they are able to break them apart. 
This is possible because resin molecules are smaller than asphaltene molecules and are able to enter and disturb the aggregate structure.

There is, however, a question left unaddressed by our molecular dynamic study. Experimental results on bitumen (see especially
the recent reviews~[\onlinecite{mullins2010},\onlinecite{mullins2011}]), are in agreement
with the asphaltene structure following the Yen-Mullins model.
According to this model asphaltene molecules tend to stack into nanoaggregates containing less than ten molecules. The aggregates
then gather into clusters approximately 6 nm long. Even though our simulations are able to probe the nanoaggregate dynamics,
they are unable to describe clusters because only small systems can be computed. The effect of the cluster dynamics
on bitumen aging is presumably of great importance and can be investigated with molecular simulations where docosane
molecules are included as an implicit solvent. This type of simulations will be the subject of a later work.

\acknowledgements
This work is sponsored by the Danish Council for Strategic Research as
part of the Cooee project.  The centre for viscous liquid dynamics
``Glass and Time'' is sponsored by the Danish National Research
Foundation (DNRF).

\end{document}